# Large Language Models in the Loop: A Stability- and Network-Aware Survey in Networked Control, Cyber-Physical, and Multi-Agent Systems

Haiping Du [a], Linping Chan [a,*]

[a] *School of Engineering, University of Wollongong, Wollongong NSW 2522 Australia*

*Corresponding author. Email address: linpingc@uow.edu.au.

**Abstract:** Modern networked control systems (NCSs), cyber-physical systems (CPSs), and complex multi-agent network systems (CNSs) increasingly rely on large language models (LLMs) for high-level decision-making. However, the slow, stochastic nature of LLMs directly conflicts with the strict stability and safety guarantees required by these physical systems. This survey presents a unified analysis of how LLMs can be admitted into the control loop of NCS, CPS, and CNS without compromising closed-loop guarantees. We organize this around a core principle: the LLM operates as a slow supervisor adjusting high-level goals and constraints, while a fast, certified inner loop maintains physical stability. Under this framework, LLM integration maps directly to classical networked control challenges, where inference latency acts as delay, API failures as packet dropouts, tokenization as quantization, and hallucinations as bounded disturbances. We assess current developments across all these three domains, highlighting that rising model capabilities are frequently accompanied by a drop in formal safety assurances. Finally, we propose concrete future research directions, identifying the widespread lack of formal stability proofs as the field's central open problem.

# 1. Introduction

## 1.1 Background and Motivation

Modern engineered systems increasingly operate not as isolated plants but as networked dynamical systems, in which many subsystems sense, compute, communicate, and act over shared channels. Three closely related families dominate this landscape. Networked control systems (NCSs) close the feedback loop across communication channels, linking system stability directly to network phenomena such as latency, packet dropouts, and data quantization [1], [2]. Cyber-physical systems (CPSs) unite computation, networking, and physical dynamics across sectors like power grids, transportation, robotics, and medicine [3], [4]. Complex multi-agent network systems (CNSs) focus on large-scale collections of interacting agents, examining how interaction topologies dictate emerging collective phenomena like consensus and synchronization, primarily through graph Laplacian analysis and algebraic connectivity [5]. Despite their domain-specific terminologies, all three families share the same foundational mathematical core of coupled dynamics communicating over uncertain networks. This survey analyzes the stability and safety of LLM-in-the-loop operation across all three of these system families under a single unified framework.

In parallel, artificial intelligence has evolved from a supplementary module into a primary architectural component within control architectures. The most ultimate shift comes from the emergence of large language models (LLMs), which demonstrate key capabilities in task planning, logical reasoning, and programmatic code generation [6], [7]. Distinct from conventional learning-based methods, LLMs introduce semantic intelligence by breaking down natural language objectives into discrete executable steps, compiling instructions into system code, interpreting multimodal context, and facilitating multi-agent coordination using human language. Early robotic applications highlighted this potential by grounding high-level text commands into actionable execution plans [8], deploying vision-language-action (VLA) models to convert internet-scale data into motor actions [9], [10], and automatically synthesizing verified logic for industrial controllers [11]. These achievements mark a fundamental evolution, moving from traditional numerical estimation toward high-level semantic reasoning and coordination integrated above classical control loops.

## 1.2 Problem Statement and Scope

This structural shift introduces a fundamental tension. Networked control, cyber-physical, and multi-agent systems routinely operate in safety-critical environments where decades of engineering research have established rigorous guarantees, including Lyapunov stability, input-to-state stability (ISS), synchronization criteria, and control barrier functions (CBFs) [5], [12]. On the other hand, LLMs function stochastically, exhibit high inference latency, and generate plausible yet erroneous outputs. They possess no native stability properties and cannot satisfy the sub-millisecond real-time demands of inner execution loops [7], [11]. Integrating an LLM into such loops directly threatens the core guarantees these systems rely on. Consequently, this survey addresses one central question: Under what precise conditions can a large language model join a networked control loop without compromising the underlying system's stability and safety?

We address this question through a unifying positioning principle found across recent research. LLMs achieve their greatest utility when operating as context-aware supervisors that adapt control targets, communication strategies, and multi-agent coordination over longer time horizons. Meanwhile, classical or learning-based controllers maintain real-time physical stability and safety guarantees. In this architecture, the LLM never replaces the lower-level controller. Instead, it modulates reference values, constraint sets, event-triggering conditions, and interaction topologies, with all outputs passing through runtime safety filters before execution. This separation is grounded in what can be rigorously proven under existing frameworks rather than a fundamental limitation. Section 6.3 discusses the theoretical advances needed to admit LLMs directly into real-time control loops.

We set clear limits to keep this survey focused and thorough. First, CNS refer only to engineered multi-agent systems studied with network tools like Laplacian graphs, consensus, and synchronization, excluding human or natural networks. Second, we include multimodal and embodied foundation models, such as VLA models [9], [10], because text-only models rarely move physical hardware on their own. Third, standard AI, reinforcement learning, and basic control methods are included only as background to highlight what is unique about LLMs. This focused scope allows us to review both real-time uses and offline tasks, such as generating code, setting reward functions, and writing system rules.

## 1.3 Related Surveys and Positioning

Several recent surveys examine LLMs in adjacent settings, but they leave the specific intersection targeted in this paper largely open. Closely related domain-specific surveys cover individual slices of this landscape. Within CPSs, Xu et al. [13] categorize the roles of LLMs as either assistants (handling perception and interaction) or central units (handling reasoning and planning) across robotics, autonomous vehicles, industrial applications, experimental platforms, and smart homes. They highlight safety, runtime verification, and deployment as primary open challenges. However, their taxonomy is organized by application domain rather than control-theoretic guarantees, and it omits the network dimension entirely. In autonomous driving,

Zhu et al. [14] survey LLMs across modular perception, prediction, planning, and control pipelines, as well as end-to-end paradigms. While they carefully account for hallucinations, latency, and safety, their focus remains limited to a single application vertical. On the networking side, Long et al. [15] review LLM-based network operations and performance optimization, specifically addressing monitoring, fault diagnosis, and resource scheduling. Similarly, Liu et al. [16] outline a six-stage LLM-for-networking workflow consisting of task definition, data representation, prompt engineering, model evolution, tool integration, and validation. Both of these studies prioritize communication networks over closed-loop control. Looking at control-theoretic foundations, K. Nosrati et al. [17] explore applying control theory to LLMs, emphasizing stability, controllability, observability, robustness, and safe deployment; however, they do not address network-related control issues. Likewise, L. Ouhib [18] reviews how LLMs assist in control system design across prompt-engineering, tool-use, multi-agent, and hybrid control-theoretic approaches. This work also skips network-related control topics. Additional general literature catalogs LLMs for multi-robot systems [19], LLMs for power systems [20], and the overall capabilities and limitations of LLMs [7].

None of these existing works offers the precise combination presented here: evaluating LLMs directly inside the loop of networked dynamical systems, analyzing them through the lens of formal stability and safety guarantees, and unifying the discussion across NCS, CPS, and CNS. In fact, existing domain surveys serve as complementary evidence for this research gap. Each survey confirms that LLMs are entering safety-critical loops within its respective vertical, and each independently highlights the exact same missing component: a bounded characterization of LLM error accompanied by a formal stability analysis built upon that bound. Table 1 summarizes this positioning along three axes vital to control practitioners: whether the survey is LLM-specific, whether it accounts for the network dimension, and whether stability and safety are analyzed as first-class guarantees rather than merely discussed.

**Table 1. Positioning relative to existing surveys.**

| Survey (domain) | LLM-specific | Network dimension | Stability / safety central |
|---|---|---|---|
| LLM-enabled CPSs [13] | Yes | No | Discussed, not analyzed |
| LLM-powered autonomous driving [14] | Yes | No | Discussed, not analyzed |
| LLM-based network operations [15] | Yes | Networking | No |
| LLMs-for-networking workflow [16] | Yes | Networking | No |
| LLMs and control theory, general [17] | Yes | No | Partial |
| LLMs in control-system design [18] | Yes | Partial | No |
| LLMs for multi-robot systems [19] | Yes | Domain-specific | No |
| LLMs for power / energy systems [20] | Yes | Domain-specific | No |
| *This survey (NCS + CPS + CNS)* | *Yes* | *Yes (unified)* | *Yes (central)* |

## 1.4 Contributions

To address these limitations, this paper proposes a unified closed-loop architecture that explicitly integrates LLM supervisory reasoning across NCS, CPS, and CNS. As illustrated in Figure 1, the proposed framework establishes a multi-layered interaction pipeline where the high-level cognitive and decision-making capabilities of the LLM act alongside certificate-based safety filters to guide lower-level physical and networked control loops.

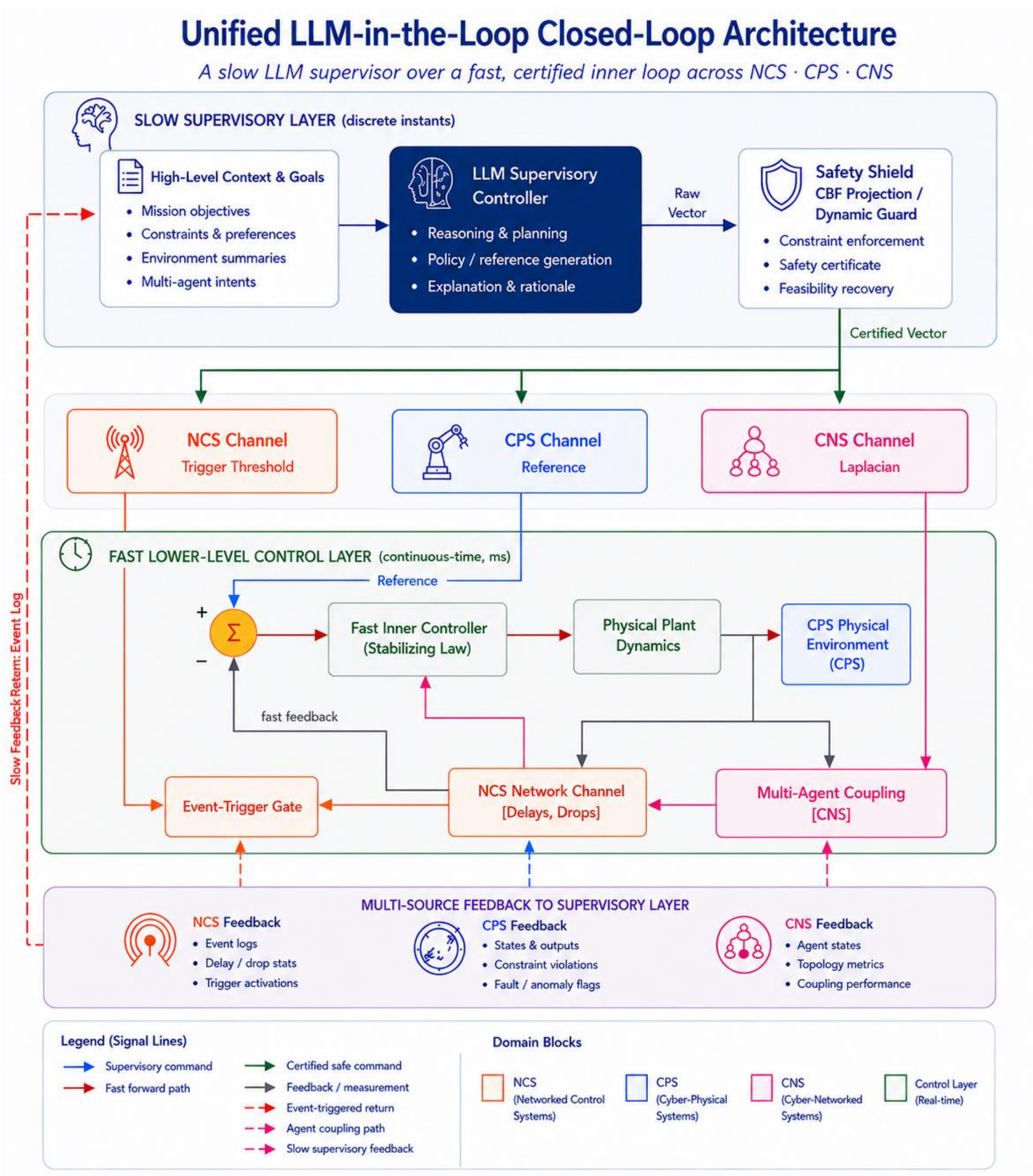

Figure 1. Overview of the proposed unified closed-loop framework integrating LLMs across NCS, CPS, and CNS domains.

In this architecture, signal flows proceed dynamically from high-level contextual instructions down to domain-specific actuation channels. Feedback paths, encompassing physical state measurements, networked delay parameters, and agent interaction states, are continuously fed back to both the local control layer and the supervisory layer. By decoupling high-level semantic planning from real-time safety enforcement, the framework guarantees system stability and bounded disturbance rejection while leveraging the adaptive reasoning of LLMs. The specific contributions are listed as follows:

1. **Unified Networked Model.** We present a joint dynamical framework for NCS, CPS, and CNS that embeds the LLM as a slow supervisor over a fast control law. Through a stability and safety lens, the LLM acts as a classical networked control element: inference latency maps to time delay, API timeouts to packet dropouts, and token decoding to signal quantization. Non-standard imperfections, such as hallucinations modeled as bounded disturbances and inherent stochasticity, are treated separately in Section 2.

2. **Classification Framework.** We organize the literature across five key dimensions: intelligence level, verification strength, network awareness, loop role, and timescale. From this, we classify existing methods into design-time, supervisory, and LLM-dominant categories.
3. **Per-System Synthesis.** We review real-world LLM deployments in NCS (communication-aware, event-triggered control), CPS (hierarchical planning, constraint generation, fault reconfiguration) and CNS (topology adaptation, coordinated control). For each domain, we assess stability status, outline future research, and analyze how LLM decisions propagate through network dynamics.
4. **Safety Toolbox and Open Problems.** We compile core safety mechanisms, including runtime safety filters, CBFs, conformal prediction, and verification certificates, into a unified perturbation-centric view. Synthesizing challenges from driving, networking, and CPS, we identify the primary theoretical gap: the lack of Lyapunov-based stability proofs for LLM-in-the-loop systems.

### 1.5 Methodology and Organization

We conducted a structured review of literature published primarily (87%) between 2021 and 2026 across control, robotics, machine learning, and systems venues. Our inclusion criteria focused on LLMs (or LLM-derived multimodal models) interacting with networked, cyber-physical, or multi-agent control problems. We stress that engineering effective LLM-in-the-loop control is inherently a multidisciplinary effort, drawing together control theory, machine learning, networking and communications, formal verification, and robotics. For this reason we make no claim of exhaustiveness. Instead, the survey highlights the developments we consider most consequential for embedding LLMs within closed control loops without forfeiting stability and safety guarantees. By unifying these strands across NCSs, CPSs, and CNSs, it is intended to give the reader a coherent and comprehensive vantage point on the LLM in the control loop, together with a principled entry point for examining individual techniques in greater depth. The remainder of this paper is structured as follows. Section 2 develops the unified model and the core LLM-in-the-loop formulation. Section 3 reviews the foundational concepts required to establish theoretical guarantees. Section 4 presents our framework and taxonomy. Section 5 surveys LLM methods across NCS, CPS, and CNS from a network-level perspective. Finally, Section 6 highlights open challenges, identifies future research directions, and concludes the paper.

## 2. A Unified Networked Model

### 2.1 General Networked Dynamical Model

We begin by fixing a single model that specializes to NCS, CPS, and CNS, so that the role of the LLM can be stated once and interpreted in each domain. Consider $N$ subsystems indexed by $i \in V = \{1, \dots, N\}$, interconnected over a graph $G = (V, E)$ with weighted adjacency entries $a_{ij} \geq 0$ and neighbour sets $N_i$. Each subsystem carries a state $x_i$ and input $u_i$, and evolves as

$$\dot{x}_i = f_i(x_i) + g_i(x_i)u_i + \sum_{j \in N_i} a_{ij} h_{ij}(x_i, x_j) + d_i, \tag{1}$$

where $f_i$ is the drift, $g_i$ the input map, $h_{ij}$ the interconnection between $i$ and neighbor $j$, and $d_i$ a disturbance collecting modeling error and exogenous inputs. Stacking the states gives the compact form

$$\dot{X} = F(x) + G(X)u + (L \otimes H)(X) + d, \tag{2}$$

where $L$ is the graph Laplacian, $H$ the coupling operator, and $\otimes$ the Kronecker product; for the diffusive couplings common in consensus and synchronization the interconnection term reduces exactly to a Laplacian acting on the state [5]. Sensing and actuation are mediated by the network: outputs are sampled and transmitted subject to delay and loss, and each local controller acts on networked estimates $\hat{x}_j$ of its neighbors rather than their true states. In the absence of an LLM the closed loop is the classical networked feedback law $u_i = \kappa_i(x_i, \hat{x}_j)$, whose stability under sampling, delay, and dropout is the subject of a mature literature [1], [2].

## 2.2 The LLM as a System Component

An LLM enters this model as a parameter generator for the control law rather than a direct input term. Let $D_i(t)$ denote the data available to subsystem $i$, encompassing measurement history, event logs, and task descriptions. In addition, let $c_i(t)$ represent a context that can include natural-language goals alongside prior interactions. The LLM is a map $\Lambda$ returning a vector of supervisory parameters

$$\theta_i(t_k) = \Lambda_i^{LLM}\big( D_i(t_k), c_i(t_k)\big), \tag{3}$$

evaluated at invocation instants $\{t_k\}$. Because inference is expensive, these instants are sparse and in general aperiodic, separated by at least the inference latency, $t_{k+1} - t_k \geq \tau_{LLM}$. The parameters are held constant between invocations (a zero-order hold) and enter the fast loop through a parameterized control law

$$u_i(t) = \kappa_i\Big( x_i(t), \hat{x}_j(t);\ \theta_i(t_k)\Big),\ \ t \in [\, t_k, t_{k+1}\,). \tag{4}$$

The result is a two-timescale, switched system: a continuous inner loop at millisecond rates under $\kappa$, and a discrete outer loop in which the LLM re-selects $\theta$ at second-to-minute rates. The parameters $\theta_i$ may encode references, constraints, triggering thresholds, controller gains, mode indices, or the coupling weights $a_{ij}$ when the LLM directly modifies the network structure. The parameter actually in force represents a delayed, quantized, and potentially dropped version of the LLM's raw output. Denoting $\hat{\theta}(t)$ as the parameter value applied at time $t$ and $t_\mathrm{k}$ as the most recent successful invocation instant, the applied parameter satisfies $\hat{\theta}(t) = q(\theta(t_\mathrm{k}))$, where $q(\cdot)$decodes discrete token sequences onto the valid parameter set $\Theta$. Under a failed or timed-out call modeled as a packet dropout, no parameter update occurs and the zero-order hold simply retains the previous value $\hat{\theta}$. Furthermore, because model inference consumes a finite latency $\tau_{LLM}$, the output parameter $\theta(t_\mathrm{k})$ is computed using state data gathered at time $t_\mathrm{k} - \tau_{LLM}$, introducing a delayed observation. These phenomena correspond directly to the time delay, packet dropout, and signal quantization introduced in Section 2.3.

This supervisory formulation directly aligns with how applied literature characterizes the role of the LLM. Across autonomous driving and CPSs, the LLM is consistently modeled as a high-level reasoning and planning layer, acting as a supervisor or "brain" above a low-level controller rather than as an inner-loop policy [13], [14]. Similarly, in networking contexts, it serves as an operator that adjusts configuration parameters above the real-time data plane [15], [16]. The mathematical model above formalizes this supervisory role and explicitly isolates the exact conditions under which it remains safe.

Three properties distinguish $\Lambda$ from a conventional supervisory map and must be modeled explicitly. First, it is stochastic and generally non-unique: identical prompts can yield different outputs, so $\theta_i$ is drawn from a distribution conditioned on $(D_i, c_i)$ rather than computed deterministically. Second, its output space is discrete: parameters are decoded from tokens and take values in a quantized set $\Theta$. Third, it is fallible: the returned parameters may deviate from the optimal supervisory choice. We formalize this property by decomposing the parameter vector as

$$\theta_i(t_k) = \theta_i^*(t_k) + \delta_i^{LLM}(t_k), \tag{5}$$

where $\theta_i^*$ is the ideal supervisory value and $\delta_i^{LLM}$ is an error term acting as the mathematical stand-in for hallucination. This error is assumed to be bounded in probability according to

$$P\left( \left\| \delta_i^{LLM} \right\| > \varepsilon \right) \leq \beta(\varepsilon), \quad (6)$$

for some tail function $\beta$ that may exhibit heavy-tailed behavior. This formulation provides a direct bridge to stability analysis. Because the control law depends on $\theta$, an error $\delta_i^{LLM}$ induces a perturbation $\Delta\kappa_i$ to the applied input; if $\kappa_i$ is Lipschitz continuous in its parameters, then

$$\|\Delta\kappa_i\| \leq L_\kappa \left\| \delta_i^{LLM} \right\|. \quad (7)$$

Consequently, a bounded LLM error translates to a bounded input perturbation. The closed-loop dynamics can subsequently be analyzed as a nominal, $\theta^*$-parameterized system driven by the perturbation $\Delta\kappa_i$, matching the exact canonical setting for ISS.

**Remark 1:** Based on the analysis in this subsection, a stability template for the LLM-in-the-loop system is given as follows: If the nominal $\theta^*$-parameterized loop is ISS with respect to input perturbations and the LLM error is bounded in the sense of (6), then the LLM-in-the-loop system inherits a practical (or stochastic) stability guarantee whose ultimate bound scales with the LLM error bound and the inter-invocation interval. This is stated as a design template, not a theorem: each surveyed method instantiates it under its own assumptions on $\kappa, \beta$, and $\{t_k\}$.

## 2.3 The LLM in the Loop as a Networked-Control Problem

The characterization above exposes a core structural insight that organizes much of this survey: the LLM does not merely sit beside a networked control loop, it introduces a second networked loop with familiar pathologies. Under this survey's stability-and-safety lens, its latency behaves as a delay, an API failure or timeout as a dropout, and its numeric token outputs as a quantization. Hallucination, by contrast, corrupts rather than erases information, meaning it is modeled as the bounded disturbance $\delta^{LLM}$ from Section 2.2 rather than a dropout. Table 2 details this correspondence explicitly and highlights the analytical tools enabled by each analogy. A key practical consequence is that existing control machinery remains directly applicable. Delay-dependent Lyapunov–Krasovskii functionals bound the impact of inference latency, while Bernoulli or Markov dropout models capture invocation failures and timeouts under which the zero-order hold retains the previous parameter $\theta$. Furthermore, standard quantizer analyses apply to the discrete set $\Theta$; and event- or self-triggered control theory governs the choice of triggering intervals $\{t_k\}$, addressing the fundamental question of when invoking an LLM call is justified [1], [2].

**Table 2. The LLM-in-the-loop / NCS correspondence.**

| LLM artifact | NCS analogue | Transferable analytical tool |
| --- | --- | --- |
| Inference latency $\tau_{LLM}$ | sensor/actuator delay | delay-dependent Lyapunov–Krasovskii functionals |
| API failure / timeout | packet dropout (hold last $\theta$) | Bernoulli / Markov dropout models |
| Hallucination (plausible but wrong output) | corrupted / accepted measurement — a bounded disturbance, not an erasure | robust / ISS analysis; disturbance $\delta^{LLM}$ |
| Tokenized output $\theta \in \Theta$ | signal quantization | quantized-control analysis |
| Aperiodic invocation $\{t_k\}$ | event-/self-triggered sampling | event-triggered stability |
| Rate limits / bandwidth | communication constraints | scheduling and co-design |

Building on this correspondence, we now interpret the unified model across each domain, mapping the context-specific meanings of the state, physical coupling, disturbances, and parameter vector as summarized in Table 3. In NCS, where the network acts as a communication channel, $\theta$ most naturally encodes triggering thresholds and scheduling decisions. This structure keeps the LLM strictly above the real-time control loop, aligning with standard practices for LLM-based network operation and configuration [15], [16]. In CPS, where the coupling represents the physical plant, $\theta$ captures reference trajectories, operational constraints, and mode or reconfiguration decisions generated through planning and reasoning. Here, the LLM fulfills its most natural and mature role by functioning as a high-level "brain" positioned above a low-level controller [8], [13], [14]. In CNS, the LLM directly acts upon the network itself. In this context, $\theta$ parameterizes the coupling weights $a_{ij}$, the topology, or the assignment of roles among agents, meaning the LLM literally reshapes the Laplacian matrix that governs system consensus and synchronization [5], [13], [19]. Across all three domains, theoretical guarantees are enforced at the control layer. Consequently, the admissibility of the LLM depends entirely on whether its slow, fallible, and quantized modulation leaves those underlying safety and stability guarantees intact, which forms the central question explored throughout the remainder of this survey.

**Table 3. The unified model interpreted in each domain.**

| Model object | NCS | CPS | CNS |
|---|---|---|---|
| State $x_i$ | local plant state | physical + cyber state | agent state |
| Coupling ($a_{ij}$, H) | communication graph | physical interconnection | interaction topology |
| Disturbance $d_i$ | delay / packet loss | environment / faults | noise / stochasticity |
| LLM parameter $\theta_i$ | triggering, scheduling | references, constraints, mode | weights $a_{ij}$, roles, topology |
| Guarantee enforced by | networked feedback | inner controller + safety filter | consensus / sync protocol |

# 3. Analytical Foundations

This section is deliberately lean, collecting only the essential tools required to formulate guarantees for LLM-in-the-loop systems and to contrast LLMs with classical learning-based control. It serves as targeted background for the framework presented in Section 4, rather than a standalone survey of control theory or machine learning.

## 3.1 Stability and Safety Tools

The model in Section 2 treats the LLM's error as an exogenous disturbance and its invocation as a form of aperiodic sampling. The four tools detailed below provide the mathematical foundation for converting that formulation into formal statements regarding stability and safety. Table 4 summarizes these tools alongside their specific roles within this survey.

**Table 4. Foundational stability and safety tools.**

| Tool | What it provides | Role in this survey |
|---|---|---|
| Lyapunov stability & ISS [21], [22] | boundedness of the state under bounded exogenous inputs | absorbs the LLM error $\delta^{LLM}$ as a disturbance (Section 2) |
| Consensus & synchronization via the graph Laplacian [5], [23] | agreement of coupled agents, set by algebraic connectivity | governs the collective behaviour an LLM reshapes (Section 5.3) |
| Event- and self-triggered control [24] | stability under aperiodic, threshold-based sampling | models the sparse LLM invocation instants $\{t_k\}$ (Section 2) |
| CBFs & safety filters [12] | forward-invariance of a designated safe set | filters LLM-generated commands before actuation (Section 4.3) |

### 3.1.1 Lyapunov stability and ISS

Lyapunov's direct method serves as the core foundation for nonlinear stability analysis [21]. Its disturbance-aware refinement, ISS, certifies that the state stays bounded by a class-$\mathcal{KK}$ decay of the initial condition plus a class-$\mathcal{K}$ gain on the magnitude of an exogenous input [22]. This property offers the natural framework for modeling the LLM as a disturbance. Specifically, if the nominal, ideally parameterized loop is ISS and the LLM parameter error remains bounded per Section 2, the perturbed loop inherits a bounded ultimate error proportional to that error bound. Formally, a system $\dot{x} = f(x, d)$ is ISS if there exist a class-$\mathcal{KL}$ function $\beta$ and a class-$\mathcal{K}$ function $\gamma$ such that the state trajectory satisfies $\|x(t)\| \leq \beta(\|x(0)\|, t) + \gamma(\sup_{\tau \leq t}\|d(\tau)\|)$ for all $t \geq 0$. Equivalently, an ISS-Lyapunov function $V$ satisfies $\dot{V}(x, d) \leq -\alpha(\|x\|) + \gamma(\|d\|)$, for a class- $\mathcal{K}_\infty$ function $\alpha$. In the absence of disturbances where $d = 0$, this bound reduces directly to the standard asymptotic-stability condition $\dot{V}(x) < -\alpha(\|x\|)$ [21], [22].

### 3.1.2 Network stability: consensus and synchronization

For multi-agent and complex-networked systems, the primary notion of stability is agreement or synchronization among coupled subsystems. The rate of convergence for these systems is governed by the spectrum of the graph Laplacian, with a particular dependence on its algebraic connectivity [5], [23]. These underlying spectral conditions dictate whether an LLM aids or degrades system performance when it modifies coupling weights or alters graph topology in Section 5.3. Concretely, the linear consensus protocol $\dot{x} = -Lx$ drives all agents to agreement at a rate set by the algebraic connectivity $\lambda_2(L)$, which corresponds to the second-smallest eigenvalue of the graph Laplacian $L$ [5], [23].

### 3.1.3 Event- and self-triggered control

Because an LLM is invoked at sparse and generally aperiodic time instances, the corresponding sampling dynamics are governed by event-triggered and self-triggered control theory. This paradigm executes sensing and actuation tasks when the system state crosses a predefined threshold instead of operating on a fixed periodic schedule, generating the precise aperiodic invocation sequence introduced in Section 2 [24]. This behavior connects directly to classical networked control literature regarding time delays and packet loss, from which this survey draws its core analogies [1], [2]. A representative self-triggered formulation defines the next execution instant as $t_{k+1} = \inf\{\, t > t_k |\, \|e(t)\| \geq \sigma\|x(t)\|\,\}$, with measurement error $e(t) = x(t_k) - x(t)$ and threshold $\sigma > 0$. The control input remains held constant between discrete triggering events, ensuring that sensing and expensive LLM computations occur only when the state drifts beyond tolerable bounds [24].

### 3.1.4 Safety certificates: CBFs

CBFs ensure that a designated safe set remains forward-invariant. When combined with a nominal controller in the form of a safety filter, a CBF minimally modifies a commanded input to guarantee the system state never leaves the safe region [12]. This safety filter serves as the core mechanism for overriding or adjusting LLM-generated commands prior to actuation in Section 4.3. Formally, for a safe set $\mathcal{C} = \{x \,|h(x) \geq 0\}$, a continuously differentiable function $h$ is a control barrier function if there exists an extended class- $\mathcal{K}$ function $\alpha$ satisfying $\sup_u[L_f\, h(x) + L_g\, h(x)u] \geq -\alpha(h(x))$ for all $x \in \mathcal{C}$. The safety filter then solves the optimization problem $u = \operatorname{argmin}_{\hat{u}} \|\hat{u} - u_{\text{nom}}\|^2$ subject to $L_f\, h(x) + L_g\, h(x)\hat{u} \geq -\alpha(h(x))$, which minimally corrects a nominal input $u_{\text{nom}}$ to keep $\mathcal{C}$ forward-invariant [12].

## 3.2 The LLM Perspective

Classical learning-based control embeds a learned function directly inside the loop and evaluates it using traditional control-theoretic tools. Adaptive and neural-network control establish closed-loop stability under parametric and functional uncertainty [26]. Reinforcement learning optimizes control policies through environmental interaction [27], with safe reinforcement learning incorporating explicit constraints or risk metrics [28]. Additionally, learning-based model predictive control (MPC) integrates learned models or learned controller parameterizations while strictly preserving state and input constraints [29]. LLMs differ fundamentally from all of these classical approaches along three core axes that structure the remainder of this survey: they manipulate semantic representations rather than numeric signals, operate at a supervisory rather than inner-loop timescale, and lack the inherent boundedness and Lipschitz continuity assumptions that learning-based analyses rely upon. Nevertheless, the connection remains concrete. Learning-based MPC already studies how to infer optimal cost functions and operational constraints, representing the parameter vector $\theta$ that yields desirable closed-loop behavior [29]. The LLM generalizes this underlying concept to natural-language-driven semantic parameterization, albeit while relinquishing the formal guarantees that classical MPC formulations retain. Consequently, learning-based control serves as the comparative baseline against which LLM contributions are evaluated rather than a primary subject of this survey. The subsequent section constructs our taxonomy and analytical framework upon these theoretical foundations.

# 4. A Stability–Safety–Intelligence Framework

This section serves as the analytical core of the survey, providing three key components: a structural decomposition and five-dimensional taxonomy that show how an LLM integrates with a control loop and locate any method in a common analytical space (Section 4.1), three method categories defined by the LLM's operational role and the surviving guarantees (Section 4.2), and the safety mechanisms that render the perturbation view of Section 2 actionable (Section 4.3). A central thesis connects these parts: intelligence enters the system as slow supervision, while formal guarantees remain firmly anchored to the control layer. Consequently, the tighter the coupling between the LLM and the fast control loop, the weaker the theoretical guarantees that can be offered.

## 4.1 Structure and Taxonomy

Every method surveyed in this work decomposes the closed-loop architecture into three primary components: the plant, a feedback controller, and an LLM module. These components connect through four distinct interaction pathways, which are presented below in increasing order of coupling to the fast control loop. Figure 2 depicts these pathways over the certified fast loop and illustrates the inverse relationship between coupling strength and theoretical guarantees.

The first pathway is interpreter or estimator. The LLM translates heterogeneous contextual information, such as natural-language goals, system logs, or execution history, into a formal state estimate or goal representation for the controller. Sitting farthest from the fast loop, this pathway is the easiest to certify. The second is supervisor. The LLM configures the parameter vector $\theta$ introduced in Section 2, setting reference trajectories, constraints, controller gains, mode indices, triggering thresholds, or network coupling weights. A certified inner-loop controller then enforces closed-loop stability. This approach mirrors classical supervisory and hierarchical control systems, as well as learning-based MPC where parameterizations are inferred online [29], and safe reinforcement learning where supervisory layers enforce constraint satisfaction [28], [30]. The third one is planner or coordinator. The LLM performs high-level task decomposition and, in multi-agent settings, coordinates agents using natural language. Rather than generating a single control input, this pathway shapes the broader interconnections and interaction policies across agents. The last one is in-loop policy. The LLM or a derived VLA model directly emits control actions [9], [31]. Representing the tightest coupling to the plant, this pathway offers the weakest theoretical guarantees.

The primary novelty of this architecture relative to classical learning-based control lies in its interface rather than its hierarchy [26], [30]. The outer loop functions as a semantic, natural-language-driven process that is intrinsically slow, stochastic, and quantized, as formalised in Section 2. More broadly, LLMs represent a specific instance of the foundation-model paradigm currently transforming autonomous systems [32]. Recent benchmarks evaluating how effectively current models reason about control problems demonstrate that while their semantic competence is real, it remains highly uneven [33].

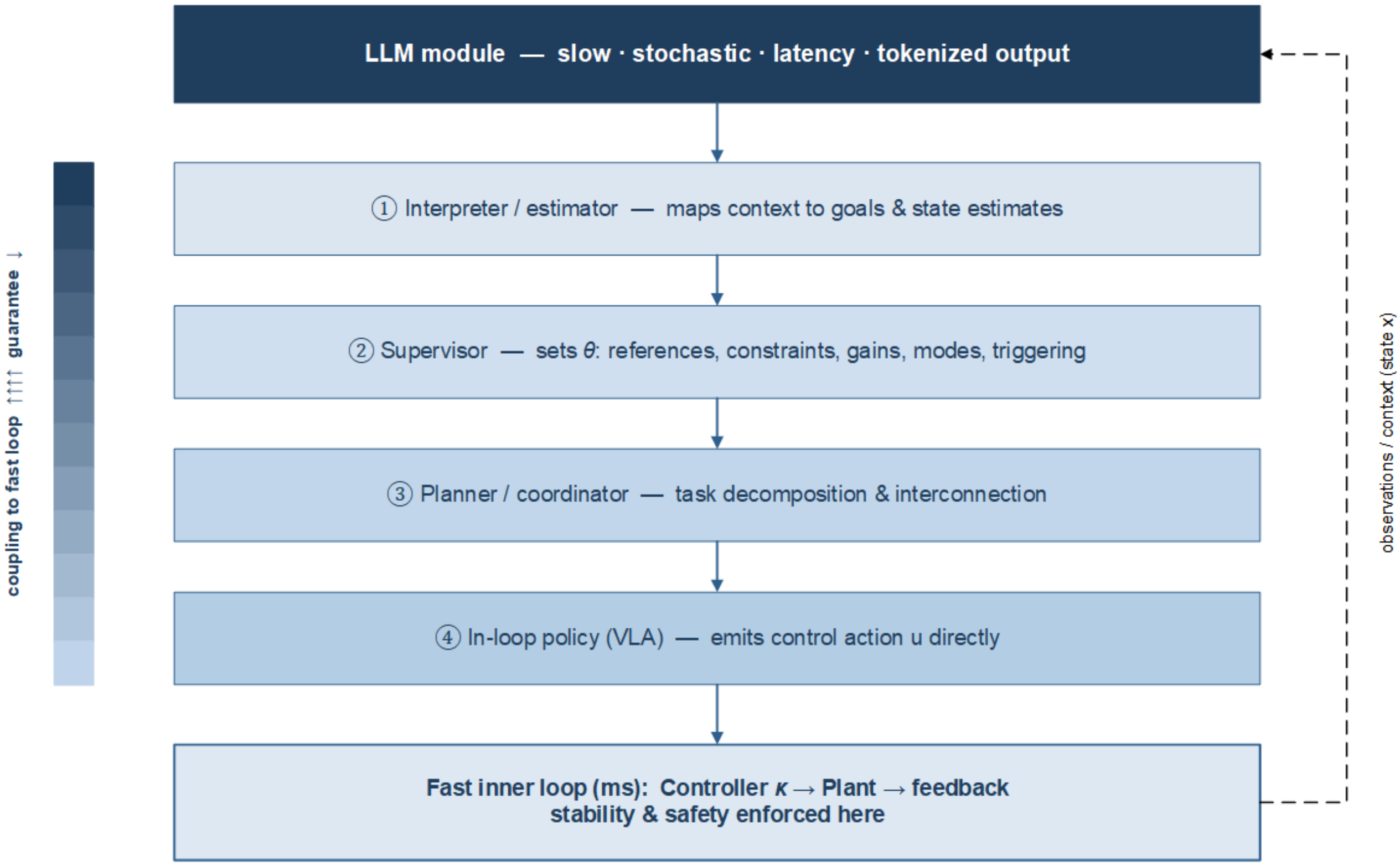

Figure 2. Integration structure of Section 4.1: the four pathways by which an LLM enters a certified control loop

We position every LLM-in-the-loop method along a five-dimensional spectrum, as detailed in Table 5. The first three dimensions stem from classical control and networked-systems analysis. The final two dimensions are specific to LLMs, representing what we argue are the most informative criteria for categorizing this literature.

**Table 5. A five-dimensional taxonomy of LLM-in-the-loop methods, with representative works.**

| Dimension | Values (increasing autonomy / decreasing guarantee) | Representative works |
|---|---|---|
| D1 · Intelligence level | model-based → learning-based → LLM-driven | [29], [26]; [27], [30]; [7], [8] |
| D2 · Guarantee / verification | proof · runtime assurance · probabilistic · partial · none | [22], [34], [35]; [12], [36], [37], [38]; [39]; [9] |
| D3 · Network awareness | centralized · distributed · decentralized | [29]; [30]; [5], [40]-[42] |
| D4 · LLM role in the loop | design-time · supervisory · planning · in-loop policy | [11], [43]; [8]; [44], [45]; [9], [31] |
| D5 · Timescale / latency | offline · seconds–minutes · sub-second · milliseconds (infeasible) | [11]; [8]; [12] |

The core organizing axis of the taxonomy is Dimension 4, the LLM's role — the same four pathways above — along which tighter coupling to the fast loop monotonically weakens the achievable guarantees, further enforced by Dimension 5, where inference latency confines LLMs to a seconds-to-minutes scale and excludes them from millisecond feedback control [13]. Three real-world systems illustrate the resulting diagonal: LLM4PLC [11] operates at design time (D4) and offline (D5), yielding proof-level guarantees (D2); SayCan [8] acts as a supervisor-planner (D4) on a seconds timescale (D5) for runtime assurance (D2); and RT-2 [9] is an in-loop policy (D4) on a sub-second timescale (D5) with essentially no formal guarantees (D2). An LLM's role (D4) and timescale (D5) thus reliably predict its formal guarantee (D2), justifying Section 4.2's organization of the literature around Dimension 4.

## 4.2 LLM-Specific Method Categories

Section 4.1 locates methods using five independent coordinates. This section reduces that description to its decisive primary axis, the role of the LLM, transforming the taxonomy's descriptive correlations into an analytical framework. Collapsing the taxonomy along this axis yields three distinct categories defined by the operational role of the LLM and the surviving theoretical guarantees, as summarized in Table 6. Grouping literature along Dimension 4 orders methods from strongest to weakest assurance without discarding the remaining axes. Within each category, the surviving guarantee is explained by the required operational timescale, while individual methods are differentiated by their level of intelligence and network awareness. The three categories therefore summarize the five-dimensional space rather than replace it, contrasting the specific guarantees that survive in each regime as illustrated in Figure 3.

In Category A, the LLM operates offline to synthesize a system artifact that is fully verified using classical techniques prior to deployment. Exemplary applications include generating verifiably correct Programmable Logic Controller and feedback controller code [11], offloading plan generation to sound classical solvers as seen in LLM+P, which uses the LLM only to translate problems into formal representations for guaranteed planners to solve [43], and translating natural language into formal goal specifications [46] optionally coupled with counterexample-guided synthesis [47]. Because generated artifacts are thoroughly verified before interacting with the physical plant, all classical stability and safety frameworks of Section 3 apply without modification. This regime offers the strongest theoretical guarantees and represents the largest body of immediately deployable literature. Automated LLM-agent frameworks further streamline this design process, such as ControlAgent coupling LLM agents with domain expertise to synthesize and iteratively refine controllers [48], or SmartControl enabling interactive LLM-driven PID controller design [49]. Reward design represents an exceptionally active sub-field where LLMs write and optimize reward functions for reinforcement learning, as demonstrated by Eureka [50], Text2Reward [51], and earlier language-to-reward methodologies [52] matching or exceeding human-engineered reward structures. Specification translation similarly thrives as LLMs convert natural-language prompts into formal temporal-logic specifications or planning objects that are subsequently enforced by model checkers and verifiers [53], [54]. Closed-loop variants such as Agents4PLC [55] and Spec2Control [56] fold verification into the generation loop.

Category B methods use the LLM to set the parameter vector $\theta$ online while a certified inner control loop enforces physical stability, representing the optimal trade-off between performance and safety for current deployments. In a canonical setup, proposed LLM parameters are grounded in a feasibility model prior to execution, such as SayCan filtering LLM plans through a learned value function so that only affordance-feasible skills are dispatched [8], while hierarchical planners refine high-level actions using environmental feedback and self-correction mechanisms [44], [45], [57], [58]. At the control interface, the LLM adjusts controller parameters rather than directly computing control actions. This approach aligns with adaptive PID schemes, the LanguageMPC framework, the parameterization view of learning-based MPC [29], and constraint-enforcing safe reinforcement learning [28], [30], all newly driven by a natural-language supervisor. The defining characteristic of Category B is an explicit trade-off between semantic performance and formal guarantees: the LLM never directly touches the fast inner loop, and its parameter updates pass through formal filters before actuation as detailed in Section 4.3. Pre-trained LLMs are increasingly deployed directly in supervisory roles for industrial control, mapping high-level system descriptions directly into operational control decisions [59].

Category C encompasses systems where an LLM or VLA model drives actions and decisions directly with minimal native theoretical assurance. Key sub-domains include end-to-end embodied policies emitting low-level control commands [9], [31], fully autonomous agentic planners that reason, execute actions, and utilize external tools [44], [57], and multi-agent LLM architectures in which several models coordinate via natural language [40]-[42], including dialectic multi-robot collaboration [42] and generative multi-agent societies [60]. While Category C exhibits the highest degree of autonomy and task generality, it provides the weakest inherent guarantees. Consequently, it represents the primary frontier where real-time safety filtering and external verification are indispensable, concentrating most of the open research challenges outlined in Section 6. The multi-agent variant involving several LLMs coordinating through language is surveyed in [61], which catalogues its open reliability and coordination challenges, while tool integration capabilities are surveyed under tool learning with foundation models [62]. A representative example of required runtime protection in Category C is the safety chip architecture [63], where an external monitor compiles user-specified safety constraints into temporal logic and vetoes any proposed action that would violate them. The requirement that safety must be enforced by such an external shield rather than guaranteed by the policy itself highlights the fundamental lack of native guarantees in LLM-dominant control architectures.

**Remark 2:** The three categories coarsen the four interaction pathways introduced in Section 4.1. The supervisor and planner pathways both fall under supervisory modulation in Category B, the interpreter pathway is typically exercised at design time in Category A, and the in-loop-policy pathway maps directly to LLM-dominant use in Category C. Because grouping is determined by the surviving guarantee, the categories order the same underlying role axis rather than re-partitioning it.

**Table 6. The three method categories, ordered by surviving guarantee.**

| Category | LLM role | Guarantee that survives | Dominant safety mechanism | Representative works |
|---|---|---|---|---|
| A — Design-time | offline synthesis, verified after | strong (classical verification) | post-hoc verification | [11], [43], [46] |
| B — Supervisory | online $\theta$-modulation | inner-loop stability preserved | safety filter / CBF | [8], [29], [45] |
| C — LLM-dominant | in-loop planning / policy | weak / empirical | runtime shield + conformal | [9], [31], [40]–[42] |

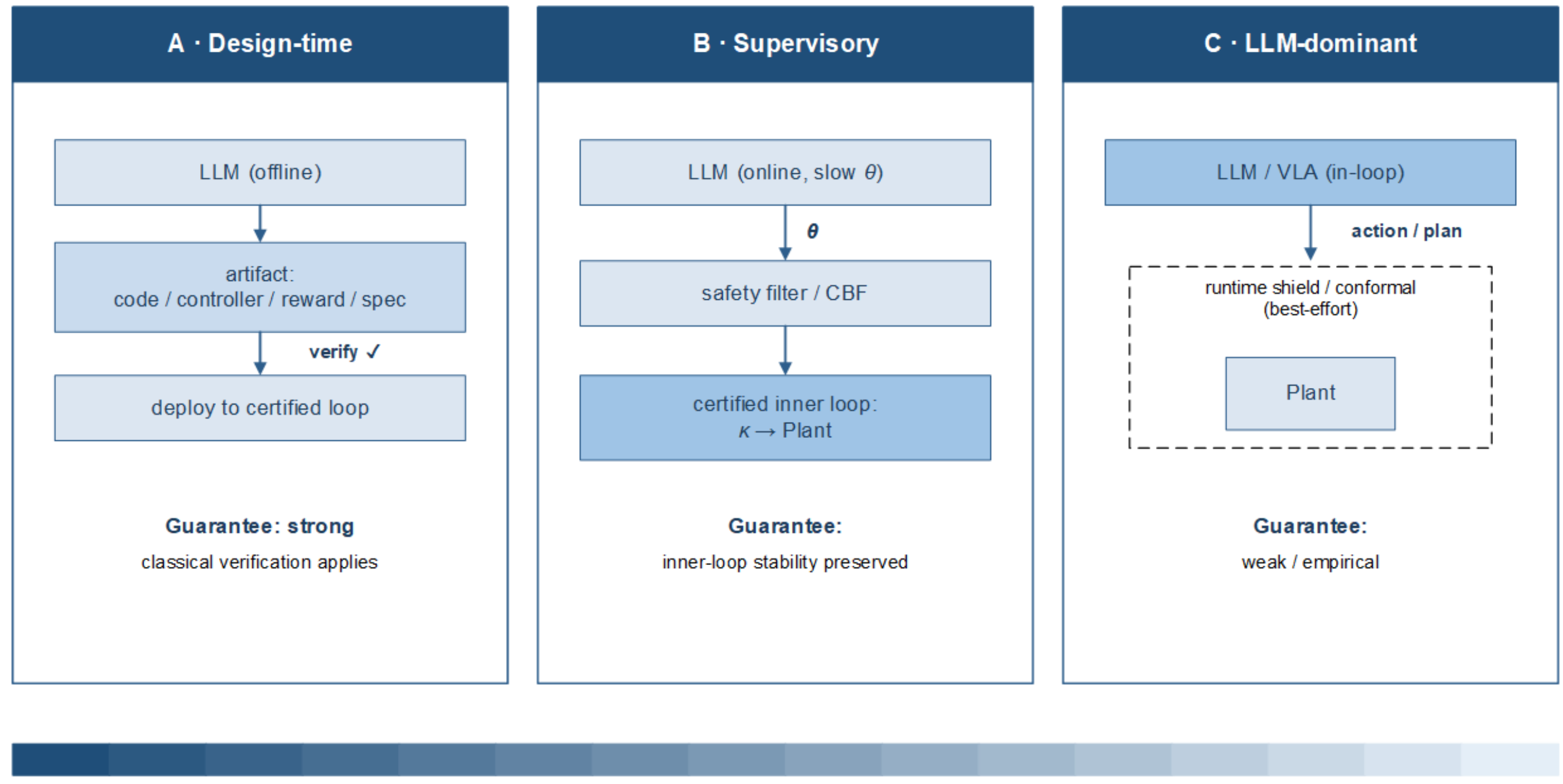

Figure 3. The three LLM-in-the-loop method categories of Section 4.2

## 4.3 Safety Mechanisms and the Perturbation View

The analytical framework concludes with the safety mechanisms that render the perturbation perspective of Section 2 operational. Recall the core theoretical template: if the nominal control loop is input-to-state stable and the LLM execution error is bounded, the integrated LLM-in-the-loop system inherits a practical stability guarantee. Every mechanism described below functions by either bounding that error or filtering it before it reaches the physical plant. The specific mechanism that becomes load-bearing is directly determined by the method's classification within Categories A, B, or C. Figure 4 illustrates this workflow, tracing the path from raw LLM error to a filtered or bounded command executing within a certified control loop.

The first mechanism is guarantee-preserving learning (nominal stability).Where the control loop itself involves learned components, closed-loop stability can be designed into the architecture. Safe model-based reinforcement learning extends Lyapunov verification techniques to synthesize policies equipped with formal stability certificates [34]. Broader safety frameworks certify learning-based control when operating in uncertain robotic environments [35], while Lyapunov-based and constrained formulations restrict reinforcement learning algorithms to designated safe sets [28], [38]. Together, these methods establish the foundational nominal ISS property required by the analytical template. The second is probabilistic guarantees on the LLM. Rather than constraining the physical plant, an alternative approach directly bounds the uncertainty of the language model. Conformal prediction provides distribution-free correctness guarantees for LLM planners, allowing the model to defer decisions when prediction uncertainty is high [39]. This technique offers the most direct pathway to establishing the probabilistic error bounds demanded by the stability template, creating a clear bridge between confident LLM outputs and certified control commands. The third one is certificate-based filtering (runtime safety). CBFs render a designated safe set forward-invariant and act as real-time safety filters that minimally modify commanded inputs to preserve set invariance [12]. Predictive safety filters wrap learning-based policies so that proposed inputs are either accepted or corrected against a certified backup controller [36], a filtering paradigm now unified across safety-critical autonomous systems [37]. In addition, Hamilton–Jacobi reachability provides safety-preserving and liveness-preserving filters backed by formal proofs [64]. To support complex dynamics where hand-crafted certificates are intractable, learned neural Lyapunov, barrier, and contraction certificates extend the approach and can be synthesized and formally verified offline [65], [66]. In the perturbation framework of Section 2, these filtering mechanisms cap the realized impact of any LLM error regardless of its origin. Finally, LLM outputs can be validated against explicit mathematical or logical constraints before execution. Neuro-symbolic, counterexample-guided synthesis pairs the LLM with a formal solver in a feedback loop until a verifiably feasible plan is constructed [47]. At runtime, execution monitors enforce temporal-logic constraints on LLM-driven agents, vetoing any proposed action that violates predefined rules [63].

These four mechanisms represent complementary instantiations of a single unifying principle: leverage the semantic capabilities of the LLM while bounding or filtering its errors before they impact the physical plant. Section 5 will demonstrate how this core principle is applied across NCS, CPS, and CNS.

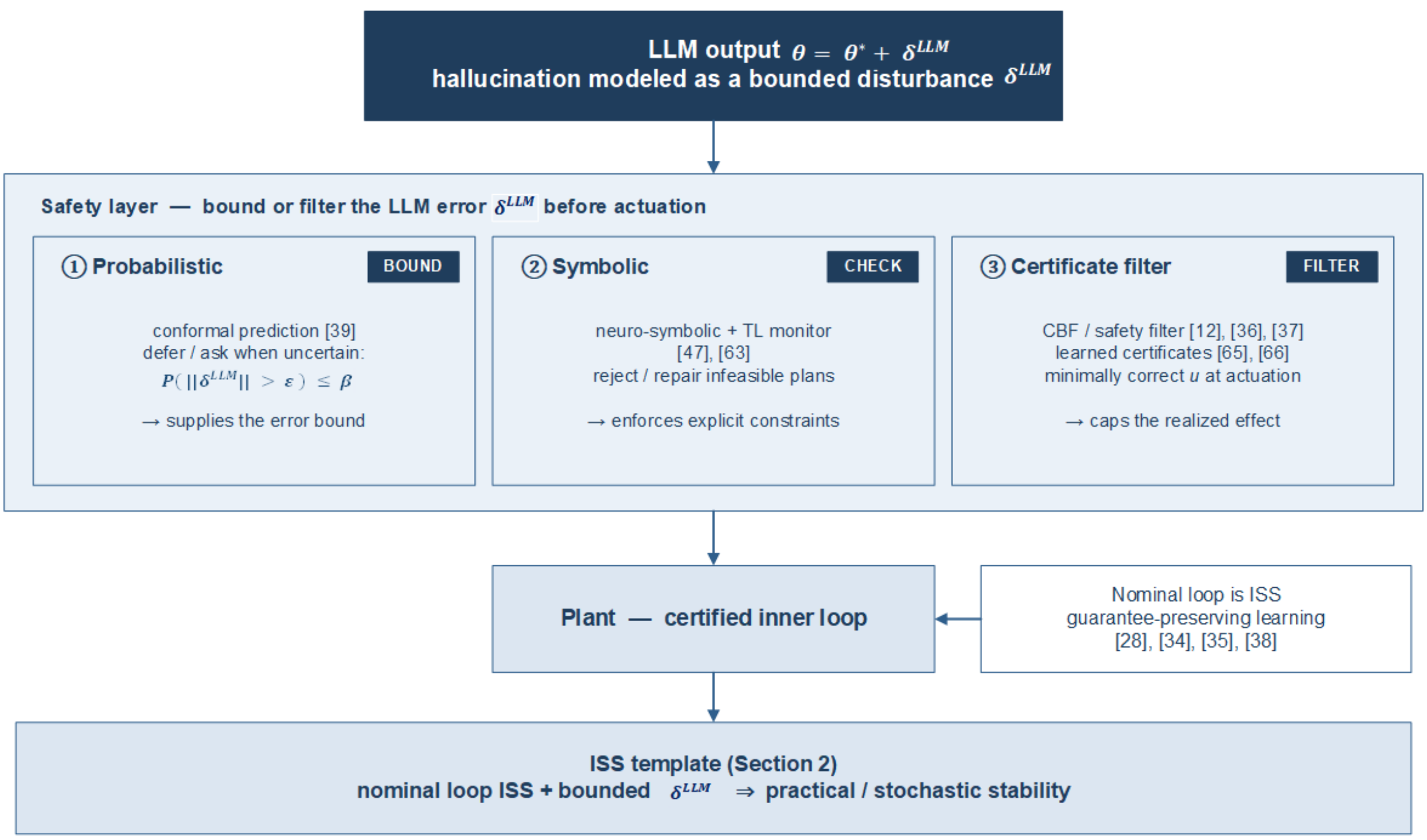

Figure 4. The logic of Section 4.3: three mechanism families bound or filter the LLM error $\delta^{LLM}$ before it reaches the plant, while guarantee-preserving learning makes the nominal loop ISS

# 5. LLMs Across NCS, CPS, and CNS

Having constructed the unified model in Section 2, established the theoretical foundations in Section 3, and detailed the analytical framework in Section 4, we now examine how LLMs are deployed across specific application domains. To maintain consistency, our survey addresses three core questions for every domain: where the LLM is positioned within the control loop, what output it is expected to generate, and to what extent classical theoretical guarantees are preserved.

## 5.1 LLMs in NCS

In NCS, the LLM operates above the real-time data plane, acting on network operation, configuration, and diagnosis rather than directly on fast feedback loops [15], [16]. Representative runtime and design-time applications span network-algorithm design [67], router-configuration synthesis and validation [68], [69], LLM-generated network-management code [70], general adaptation of LLMs to networking tasks through task-specific encoders [71], protocol fuzzing for security testing [72], intent-based network management [73], LLM-enabled reasoning over dynamic network graphs [74], optical-network alarm analysis [75], and domain-specialized models such as Mobile-LLaMA for 5G analysis [76]. In wireless and 6G infrastructure, LLM agents sit in the operation loop through perception–grounding–alignment frameworks [77], telecom-specific foundation models and surveys [78], [79], and LLM-enhanced multi-agent coordination systems [80]. Furthermore, LLMs support network monitoring and diagnosis via log- and time-series-based anomaly detection [81], [82]. This wireless and 6G expansion represents a rapid development across 2024 and 2025, positioning networking as one of the fastest-growing arenas for LLM-in-the-loop operation.

These networked implementations excel at semantic tasks such as configuration, diagnosis, and intent translation, while remaining conservative regarding loop placement by keeping the LLM strictly above the millisecond control path. This configuration reflects the supervisory role formalized in Section 2, where the LLM modulates triggering and scheduling parameters that an event-triggered mechanism subsequently enforces in real time [24]. However, surveyed NCS research remains overwhelmingly operations-oriented, with almost no literature offering formal stability analysis for the resulting closed loop, leaving ISS style treatments as an open next step. Regarding safety and security, exposure in NCS centers on the operations pipeline rather than the fast loop. A compromised or hallucinated configuration, schedule, or routing decision can misdirect multiple loops simultaneously, requiring strict validation of LLM-set parameters alongside runtime filtering mechanisms catalogued in Section 4.3, while millisecond feedback loops retain classical guarantees.

Defensively, LLMs are deployed for intrusion, DDoS, and log-based detection and mitigation [15], [83], [84], even as backdoor attacks specific to networked deployments remain a documented threat [85]. Offensively, LLM multi-agent systems can autonomously generate large libraries of industrial control system attack patterns that surpass expert-crafted baseline attacks, as demonstrated by AttackLLM [86]. Because the LLM sits above the loop, its latency and occasional invocation failures affect the system similarly to the delay and packet-dropout phenomena classically handled in NCS analysis, mapping onto delay-dependent Lyapunov–Krasovskii and Bernoulli or Markov dropout models applied to the supervisory-parameter channel rather than the sensor-actuator link.

Looking forward, LLMs are firmly established across network operation, configuration, diagnosis, and wireless management, yet they sit deliberately above the real-time data plane without formal control-theoretic guarantees. The near-term focus centers on tighter, stability-aware integration by co-designing LLM-set triggering and scheduling with event- and self-triggered mechanisms to ensure ISS or finite-time guarantees [24], [37]. Concurrently, deploying edge-resident small models shrinks invocation intervals, exemplified by self-hosted, offline edge LLMs executing control-loop reasoning locally on embedded GPU platforms [87]. Advancing this field further requires building standardized closed-loop NCS benchmarks that evaluate stability preservation under LLM-modulated scheduling rather than assessing task accuracy in isolation. Table 7 summarizes the comparison between the reviewed LLM-in-NCS approaches.

**Table 7. Comparison of LLM-in-NCS approaches**

| Approach | Works better when | Guarantee / trade-off | Evidence (type; works) |
|---|---|---|---|
| Operations, config, diagnosis, intent translation (supervisory, above the loop) | Tasks are semantic and non-real-time; the fast loop must keep its classical guarantees | Safe/conservative, but the operations pipeline itself is unanalyzed; a hallucinated config can misdirect many loops at once | Deployments and surveys at task-accuracy level, no closed-loop stability [67]-[75], [81], [82]; domain-tuned Mobile-LLaMA [76]; 6G agents [77]-[80] |
| LLM-set triggering/scheduling co-designed with event-/self-triggered control | Formal closed-loop guarantees (ISS or finite-time) are required, not just task accuracy | The only route to stability guarantees in NCS, but still an open direction | Proposed direction, thin support [24], [37] |
| Edge-resident / offline small LLMs | Latency-critical or secure operation; shorter invocation intervals needed | Shrinks the inference latency; capability cost of a smaller model | Single embedded-GPU demonstration [87] |
| Security use (defensive vs offensive) | Defensive detection (intrusion, DDoS, log analysis) is the mature win | Also enlarges the attack surface | Detection deployments [83], [84]; comparative result: AttackLLM surpasses expert-crafted baselines [86]; backdoor threat [85] |

## 5.2 LLMs in CPS

In CPS, LLMs find their most developed and natural role acting as a high-level reasoning and planning brain operating above a low-level controller [13]. Surveyed applications demonstrate this architecture across four primary domains before encountering shared stability and safety challenges.

In robotics and embodied control, the dominant architectural pattern is hierarchical. An LLM decomposes natural-language goals into feasible skills grounded in robot affordances [8], scales to large scenes via 3D scene graphs and classical path planners [88], guides decoding with grounded models [89], and reasons over skill dependencies for long-horizon tasks [90]. Recent manipulation systems extend this pattern by composing 3D value maps for zero-shot trajectories in VoxPoser [91], expressing tasks as online relational keypoint constraints in ReKep [92], closing the loop with environment feedback in Inner Monologue [93], and advancing general low-level control using flow-based VLA policies such as $\pi_0$ [94] or open imitation models such as RoboFlamingo [95]. Conversely, embodied multimodal models push perception directly into the network, as seen in PaLM-E interleaving continuous sensor data with text [31], while VLA models such as RT-2 emit actions directly, operating in the in-loop-policy regime with the weakest guarantees [9].

A complementary code-generation line enables LLMs to synthesize executable robot policies. This spans from the foundational Code-as-Policies formulation [96] to closed-loop modular synthesizers like ModuLoop that insert debugging probes and refine control code online [97]. Retrieval-augmented planners with guarded execution such as ARRC further raise plan validity while enforcing workspace and force limits at runtime [98]. Natural-language command interfaces are also validated on physical hardware through protocol-agnostic LLM-to-drone bridges exposing flight tools over the Model Context Protocol [99], as well as embedded voice-controlled assistants mapping spoken commands to a six-degree-of-freedom arm on a Raspberry Pi [100], while comprehensive surveys map this LLM contribution across the full perception-to-control navigation pipeline [101].

Autonomous driving and mobility exhibit the full spectrum of LLM operational roles. These include interpretable end-to-end driving [102], knowledge-driven decision-making [103], vision-language planning [104], multimodal alignment with behavioral planning [105], GPT-based motion planning [106], closed-loop language-guided driving [107], graph-VQA driving [108], and initial real-world physical driving experiments [109]. At the control interface, LLMs regularly modulate controller parameters $\theta$ rather than replacing the controller entirely. For instance, LanguageMPC uses high-level LLM decisions to adjust MPC parameters via a transformation matrix [110], while related systems tune adaptive PID controllers for truck platooning [111]. Recent advances integrate closed-loop motion-planning agents [112], explicit safety designs bounding LLM influence [113], split edge-cloud inference for real-time motion planning [114], and vision-language-model-guided MPC coupling scene understanding to planners in VLM-MPC [115]. This LLM-augmented MPC formulation extends beyond road vehicles to marine craft, where an LLM estimates residual (unmodeled) dynamics incorporated into the MPC prediction model for surface vessels operating under uncertain sea conditions [116].

Beyond mobility, LLMs integrate with digital-twin data to control modular production [117], act as experiment designers writing and executing scripts for chemistry and biology automation [118], [119], and orchestrate home IoT devices [120]. On physical testbeds, LLMs have been benchmarked as direct closed-loop controllers, such as greenhouse climate-control studies evaluating prompt-only, SQL-assisted, and prediction-assisted GPT-4o variants under tracking tasks with live plants [121]. Fault handling is an emerging focus, with LLMs providing explainable fault diagnosis [122], machine fault frameworks [123], and real-time anomaly detection with reactive replanning [124].

In heavy industry and infrastructure, agentic systems close the loop around PLC/DCS code generation and verification [55], [56], expanding from Structured Text to graphical languages such as Ladder Diagram and Function Block Diagram [125] while driving digital-twin production [126]. In power systems, LLM multi-agent controllers manage grid operations [127], perform in-context MPC-style distribution-voltage regulation under changing topologies [128], and automate power-converter modulation design using physics-informed agents in PE-GPT [129]. In transportation, LLMs serve as interpretable traffic-signal controllers [130]. Broader agentic frameworks automate end-to-end controller design in AgenticControl [131] or act as generalized control engineers equipped with large tool libraries across diverse plants in LLM-Agent-Controller [132], with related work pushing LLM assistance upstream to electromechanical component selection and co-design [133].

Across all these domains, the consistent picture is rising intelligence alongside falling guarantees, characterized by the LLM acting as the brain and the controller as the muscles [13]. Open control-theoretic questions regarding robustness to LLM-induced reference and constraint uncertainty, as well as stability across LLM-triggered mode switches, map directly onto the perturbation and switched-system model of Section 2, yet remain largely unanalyzed in current literature. Recent progressions from value-map manipulation to flow-based policies and open-loop to closed-loop driving agents sharpen this tension. Because CPS involves irreversible physical stakes, an erroneous or adversarial LLM output can drive actuators into constraint violations, making the bounding of LLM error the load-bearing requirement rather than merely improving average behavior. Concrete safeguards include safety-aware task planning that screens instructions prior to execution [134], semantic motion-safeguards keeping manipulation within safe sets [135], and certified runtime safety filters or CBFs that screen every command before physical actuation using shared mechanisms catalogued in Section 4.3.

Ultimately, CPS represents the most developed and varied arena for LLMs in the loop while presenting the sharpest tension between intelligence and formal guarantees as VLA models push LLMs from supervisory roles toward in-loop policies. The path forward involves wrapping LLM and VLA outputs in certified safety filters and CBFs prior to actuation [12], [36], analyzing LLM-triggered mode changes using switched-system and fault-tolerant theoretical frameworks, and pairing LLMs with observers and robust controllers so that semantic proposals are estimated and compensated downstream. Finally, addressing these challenges requires replacing today's open-loop, simulation-bound evaluations with rigorous, closed-loop, stability-aware benchmarks. Table 8 summarizes the comparison between the reviewed LLM-in-CPS approaches.

**Table 8. Comparison of LLM-in-CPS approaches**

| Approach | Works better when | Guarantee / trade-off | Evidence (type; works) |
|---|---|---|---|
| LLM modulates controller parameters (MPC/PID theta; controller unchanged) | A certified controller already exists and semantic adaptation is wanted without surrendering guarantees; safety-critical | Best guarantee/capability trade-off: controller keeps guarantees and LLM influence can be explicitly bounded | LanguageMPC [110], PID platooning [111], VLM-MPC [115], marine vessel [116]; explicit safety-bounding design [113] |
| Hierarchical planning (LLM decomposes goals into skills) | Long-horizon semantic tasks needing decomposition; robot affordances available | LLM stays supervisory, so the controller keeps guarantees (strong posture) | SayCan [8], scene graphs [88], VoxPoser [91], ReKep [92], Inner Monologue [93] (simulation + robot demos) |
| Code-as-Policies / online code synthesis | Task is expressible as code; design-time verification or closed-loop refinement is possible | Verifiable offline; runtime guards enforce workspace/force limits | Code-as-Policies [96], ModuLoop [97], ARRC [98]; physical hardware [99], [100] |
| VLA / in-loop policy (model emits actions directly) | Reactive low-level control where internet-scale pretraining transfers | Tightest coupling to the plant, so the weakest guarantees | RT-2 [9], PaLM-E [31], flow/imitation VLA [94], [95] (demonstration-level, no formal guarantees) |

| Approach | Works better when | Guarantee / trade-off | Evidence (type; works) |
|---|---|---|---|
| LLM as direct closed-loop controller | Rarely; slow plants tolerant of latency; mainly a probe of native competence | No guarantees; competence real but uneven | Comparative benchmark: greenhouse GPT-4o (prompt-only vs SQL- vs prediction-assisted) on live plants [121] |

## 5.3 LLMs in CNS

In multi-agent and complex-network systems, the object acted upon by the LLM is the network itself, including inter-agent coupling, assigned roles, and overall topology. Surveyed applications demonstrate this across a wide range of operational domains, including LLM-coordinated multi-robot systems [19], GPT-in-the-loop decision-making for self-adaptive multi-agent IoT systems [136], LLM-enabled reasoning over dynamic network graphs [74], and multi-vehicle coordination emerging from LLM decision-makers [110]. In aerial and mobile robotics, role-adaptive LLM control coordinates UAV swarms [137]. This is complemented by language-driven drone-swarm frameworks that pair an LLM waypoint generator with an explicit safety filter, such as an optimization-based corrector for synchronized choreography in SwarmGPT [138] or a real-time potential-field guard for collision-free natural-language swarm control in SkySim [139], as well as fine-tuned LLMs that guide distributed model-predictive control for decentralized UAV formations [140].

This scope extends into broader multi-agent and infrastructure coordination. Modular frameworks assemble cooperative embodied agents from LLMs [142], including embodiment-aware operating systems that assign tasks across heterogeneous robot teams in EMOS [143], planner-executor frameworks that decompose human instructions into robot-specific subgoals for quadrotor, quadruped, and manipulator fleets in COHERENT [144], and actor-critic schemes that coordinate large populations of LLM agents for scalable decision-making [145]. LLMs have also been used to resolve deadlocks in multi-robot systems, restoring progress where classical coordination stalls [146]. In transport and communication networks, collaborative-driving systems let vehicles negotiate and coordinate through language [147], [148], supported by dedicated surveys mapping multi-agent LLM autonomous driving [141], while cooperative LLM agents perform network-wide traffic-signal control [149] and LLM-enhanced multi-agent frameworks organize 6G network nodes [80]. Together, these developments extend the language-as-communication question to genuinely networked, many-agent settings, though the underlying reliability and coordination limits of such collectives remain an open concern [61].

This transition to semantic communication raises a fundamental networked-systems question with no classical analogue: when agents exchange semantic, linguistic messages rather than numerical states, what do consensus and synchronization mean over a language-valued communication graph, and what is the quantization error of a sentence? While classical foundations relying on Laplacian-spectrum conditions for agreement and synchronization [5], [23] define the target, surveyed complex-network work remains heavy on empirical LLM usage and light on rigorous stability analysis. Consequently, the field is currently best anchored to multi-robot control and distributed NCSs rather than abstract network science, exactly the scoping argued in Section 1.2. The defining recent trend is this shift from single LLMs toward multi-agent collectives, accompanied by the first concrete safety alarms, including demonstrations that LLM-controlled robots can be jailbroken into unsafe physical actions [150].

From a control-theoretic standpoint, stability and safety in these systems present two distinct collective challenges. The stability question is collective, focusing on whether consensus and synchronization survive an LLM that reshapes weights, roles, or graph topology, a property that surveyed work almost never certifies. The safety question centers on contagion. Because agents are coupled, a single compromised or hallucinating agent can propagate errors across the graph before detection occurs. Preventing this requires distributed safeguards, specifically per-agent runtime filters and network-level certificates that keep the total coupling within a safe, convergent set, drawing on the shared machinery catalogued in Section 4.3.

As a result, complex network systems represent the arena of heaviest LLM usage paired with the lightest degree of stability analysis. LLMs already actively coordinate swarms, cooperative-driving fleets, and network-wide controllers, yet almost none of this work certifies the resulting collective behavior. The open agenda is therefore largely theoretical, requiring the extension of consensus and synchronization guarantees to language-valued communication graphs, the quantification of the quantization error of semantic messages, and the design of distributed safety filters that keep many-agent coordination within a certified set. Alongside these theoretical needs lies the practical requirement for communication-efficient, scalable LLM coordination whose convergence can be rigorously analyzed as the agent count grows [5], [23], [61]. Table 9 summarizes the comparison between the reviewed LLM-in-CNS approaches.

**Table 9. Comparison of LLM-in-CNS approaches**

| Approach | Works better when | Guarantee / trade-off | Evidence (type; works) |
|---|---|---|---|
| LLM + explicit certified safety filter (waypoint generation + corrector/guard) | Safety-critical swarm/coordination where a certified filter can wrap the LLM (e.g. collision avoidance) | The only CNS pattern with a real safety guarantee today: the filter owns safety, the LLM only proposes | SwarmGPT optimization corrector [138], SkySim potential-field guard [139], fine-tuned LLM + distributed MPC [140] |
| Multi-agent coordination, topology and role adaptation (no certificate) | Flexible, language-driven coordination of heterogeneous teams; scalable decision-making | Heaviest usage but lightest analysis; contagion risk (one bad agent spreads); collective stability rarely certified | Multi-robot [19], EMOS [143], COHERENT [144], agent populations [145], cooperative driving [147], [148], traffic-signal [149], 6G [80] (demonstration-level) |
| LLM-aware consensus/synchronization over language-valued graphs | Theoretical; not yet realized in practice | The open agenda: extend consensus/synchronization and quantization to semantic messages | None yet; classical foundations [5], [23]; reliability concern [61]; jailbreak alarm [150] |

## 5.4 Network-Level Perspective

Synthesizing across the three domains reveals a cohesive network-level framework that aligns with the compact model ($L \otimes H$) introduced in Section 2. Collective behaviors such as consensus, synchronization, and their breakdown are fundamentally governed by graph topology and coupling mechanisms [5], [23]. When a LLM alters references, coupling weights, or network topology, it directly influences system stability. The surveyed literature highlights three primary network-level risks: topology-unaware decisions that are locally logical but globally contradictory, destabilizing perturbations that spread across network couplings before detection, and coordination breakdowns stemming from divergent LLM outputs among agents. To mitigate these risks, the distributed-intelligence agenda focuses on communication-efficient coordination, local versus global stability, and scalable supervision over large agent cohorts [15], [19]. Early safety-filtered swarm frameworks [138], [139] demonstrate how to maintain multi-agent coordination within certified boundaries. Central to this framework is a foundational principle: network-level guarantees must be enforced by the underlying coupling and control protocol. The slow, fallible modulations of an LLM are admitted only when they leave these guarantees intact. Looking ahead, the goal is to develop LLM-aware guarantees that treat language-driven agents as bounded, filterable perturbations on the graph, ensuring that consensus and synchronization certificates survive semantic interventions. Developments over 2024 and 2025 have reinforced this structural picture. Systems have grown increasingly capable, agentic, and distributed across domains ranging from LLM-driven control design [48] to multi-agent 6G coordination [80] and cooperative driving [147], [148].

# 6. Open Challenges, Future Directions, and Conclusion

The literature on LLMs in the control loop is young but already converging on a shared set of fundamental obstacles. Reading recent domain-specific surveys covering LLM-powered autonomous driving [14], LLM-based network operation and optimization [15], networking applications [16], and LLM-enabled CPSs [13] through the lens of our unifying framework reveals a striking consistency. Regardless of whether the physical plant is an autonomous vehicle, a communication network, or a robotic platform, identical difficulties consistently recur across disciplines. This observation reinforces the insight from Section 1.3 that these domain-specific surveys serve as complementary evidence for a single underlying theoretical gap. These recurring obstacles separate cleanly into theoretical challenges regarding whether formal guarantees can be established at all, and practical challenges regarding whether a guaranteed system can be built and deployed in real-world settings. Table 10 synthesizes these recurring difficulties along with their root causes. The subsequent subsections examine each category in turn before outlining future research directions and concluding the survey. Throughout this prospective analysis, we maintain our three-system lens by highlighting where each specific challenge bites hardest, whether in NCS, CPS, or CNS.

**Table 10. Cross-domain open challenges for LLMs in the loop, with representative sources.**

| Challenge | Manifestation across NCS / CPS / CNS | Sources |
|---|---|---|
| Hallucination as unbounded error | confident, ungrounded outputs; minimal error tolerance in safety-critical loops | [13]-[15], [39], [25], [156] |
| Continuous-signal & temporal reasoning | poor abstraction of speed/position; weak handling of deadlines and event order | [13], [33] |
| Constraint satisfaction & verification | iterative validate-and-correct still fails; NN verification infeasible at LLM scale | [13], [43], [47], [157]–[159] |
| Latency & real-time feasibility | inference delay, cloud transmission, prompt-size limits vs. inner-loop timing | [13]-[15], [114], [154] |
| Deployment & the edge | cloud–edge–device split; distillation/quantization; 5G/6G | [13], [15], [16], [76], [155] |
| Data quality & benchmarks | scarce rare-fault data; no standard benchmark; simulation-to-reality gap | [13]-[15], [33], [107], [108], [152], [153] |
| Security & privacy | adversarial inputs, poisoning, backdoors; leakage of sensitive data | [13]-[15], [150], [160]-[162] |
| Interpretability | black-box decisions impede certification and operator trust | [14], [15], [47] |

## 6.1 Theoretical Challenges

The foundational theoretical obstacles in this field reduce to a single core reality: the output error of a Large Language Model is neither small nor well-characterized, and current methodology cannot reliably bound it. This limitation creates a direct conflict with the fundamental assumption underlying the ISS template introduced in Section 2, which requires a bounded disturbance to guarantee closed-loop stability.

The first theoretical challenge is hallucination functioning as an uncharacterized disturbance within the control loop. Every surveyed domain identifies hallucination, which is defined as confident but ungrounded output, as its central operational risk [13]-[15]. In safety-critical autonomous driving, tolerance for such error is minimal [14], whereas in CPSs, hallucination can cause a controller to confidently execute an incorrect plan that violates physical constraints, such as a robotic manipulator colliding with an obstacle [13]. Within the framework of this survey, hallucination represents the disturbance term $\delta^{LLM}$ from Section 2, and the primary difficulty is that $\delta^{LLM}$ is not yet a bounded disturbance. Because an LLM remains confident even when incorrect, its true error rate cannot be read directly from raw output probabilities [13], rendering the probabilistic bounds required by the ISS template unavailable in general. Establishing such a bound, whether through conformal prediction [39], uncertainty estimation [15], or dedicated verifier models, represents the single most important open theoretical problem in the field. Beyond domain-specific surveys, hallucination is studied in its own right through dedicated taxonomies and benchmarks [25]. Although the earliest mitigation strategies attach statistical guarantees to planner outputs [39] or fold unreliable-output costs directly into the objective function [156], none currently deliver the uniform error bounds that the ISS template requires.

The second theoretical challenge stems from the weak handling of continuous signals and temporal constraints by current language models. A recurring finding across the literature is that LLMs abstract poorly from the continuous, numerical data on which control systems depend, such as speed and position, while also struggling with temporal constraints like deadlines and event ordering [13]. This phenomenon restates the timescale-and-representation mismatch established in our taxonomy, reflecting that the LLM reasons using discrete tokens over high-level semantics, whereas the physical plant evolves continuously over time. This operational mismatch explains why surveyed systems consistently isolate the LLM within a supervisory outer loop rather than placing it inside the millisecond inner loop [13], [14], ensuring that theoretical guarantees remain firmly anchored to the control layer.

The third theoretical challenge involves attempting constraint satisfaction without formal mathematical proofs. Several existing architectures attempt constraint-guaranteed planning by iteratively querying the LLM and validating its outputs against external tools, supplying counterexamples until a feasible plan is found [13]. However, the intrinsic ability of the LLM to respect constraints has not measurably improved, causing these iterative loops to frequently fail to return a valid plan within their allotted execution budget [13]. Runtime verification of the LLM itself presents an equally formidable challenge, as classical neural-network verification is computationally infeasible at LLM scale and fails to transfer cleanly to multimodal models whose inputs combine text and images and whose outputs are natural language rather than numerical vectors [13]. The direct consequence of these limitations is the pervasive gap emphasized throughout this survey, marked by the near-total absence of Lyapunov- or ISS-based analyses for LLM-in-the-loop systems [7]. In practice, the validate-and-correct research thread combines neuro-symbolic synthesis [47], temporal-logic safety chips [63], and offloading to sound classical planners [43] as pragmatic workarounds. Meanwhile, the scaling barrier of formal verification is exemplified by exact neural-network verifiers such as Reluplex [157] and Verisig [158], as well as recent methods for DNN-controlled systems [159], which fail to scale to LLM-sized architectures. Ultimately, the robust certificate machinery that currently exists for learned controllers, ranging from stability-guaranteed reinforcement learning [34] and general safety frameworks [35] to learned Lyapunov and barrier certificates [30], [65], [66], has no functional LLM-in-the-loop counterpart yet.

## 6.2 Practical Challenges

Even where a theoretical guarantee can be successfully argued, building and deploying a trustworthy system introduces a second set of practical obstacles that the surveyed domains encounter in common.

The first practical challenge involves latency and real-time feasibility constraints. The computational cost of inference makes real-time response the dominant practical constraint across systems [14], [15]. Cloud-hosted models add transmission delays and prompt size limits that complicate multi-turn consistency, while on-device deployment is often infeasible on the hardware typically carried by vehicles and robots [13]. This is the concrete reason why the inference latency parameter $\tau_{LLM}$ is large and why invocation instants must remain sparse, establishing the exact operating point that our two-timescale model was built to describe. Concrete responses are emerging through efficient inference methods that cut latency directly [154] and edge-cloud splits that offload heavy reasoning while keeping time-critical actions local [114]. To address these hardware bottlenecks, the literature increasingly converges on a cloud-edge-device split featuring heavy reasoning in the cloud alongside local pre-processing and time-critical action [13],[15]. Distillation, quantization, and small language models are repeatedly proposed to move the feasibility threshold, supported by edge computing and advanced network architectures to bring computation nearer to the physical plant [15], [16]. Domain-specialized small models such as Mobile-LLaMA for 5G analysis [76] and compact open models such as Gemma [155] make the on-device side of this architectural split practical.

The second practical challenge centers on data quality and evaluation. System performance depends heavily on the quality and diversity of training and fine-tuning data, which remains exceptionally difficult to obtain for rare faults and critical corner cases [14], [15]. Evaluation methodologies are equally unsettled, as there is currently no standardized benchmark for LLM-in-the-loop control systems. Metrics vary widely, open-loop tests suffer from causal confusion, and the simulation-to-reality gap frequently inflates reported performance [13], [14]. While robotics and autonomous driving offer partial benchmarks such as grounded instruction suites and multimodal driving datasets, most other application settings still rely on ad hoc, custom evaluation [13]. The partial benchmarks that do exist span embodied instruction following like ALFRED [152] and BEHAVIOR [153], graph visual question answering and closed-loop driving like DriveLM [108] and LMDrive [107], and control competence tests [33]. However, none of these currently score closed-loop stability under LLM-induced perturbation, which is the exact metric this survey advocates for.

The third practical challenge involves security and privacy vulnerabilities. Admitting an LLM into the control loop significantly enlarges the system attack surface because adversarial inputs, data poisoning, and backdoors can easily steer model outputs. In a physically coupled loop, a manipulated instruction directly translates into a manipulated actuation [13], [14]. Furthermore, handling sensitive operational data such as vehicle locations, driving habits, and network logs raises severe leakage risks [14], [15]. These acute concerns explain why a runtime safety filter functions as a non-negotiable architectural component rather than an optional add-on. Primary studies make this attack surface concrete by demonstrating that LLM-controlled robots can be jailbroken into executing unsafe physical actions [150], [161], and that agent memory and knowledge bases can be successfully poisoned [162], while dedicated security and privacy surveys comprehensively catalogue these vectors and their corresponding defenses [160].

The fourth practical challenge is the lack of interpretability. Black-box decisions impede the operator's ability to safely manage and certify the system. Improving interpretability through feature attribution, visualization, or knowledge-graph and neuro-symbolic structure recurs as a foundational prerequisite for trust in both the driving and network operations literature [14], [15].

## 6.3 Future Directions

The future directions that emerge from these theoretical and practical obstacles split naturally along the core organizing axis of this survey: making the LLM's contribution safer while simultaneously making the underlying guaranteed system more capable.

The first prospective direction is the development of validation environments and digital twins. A promising route to safety is wrapping the LLM in a validation environment, such as a sandbox or digital twin, that runs synchronously with the physical plant to monitor state and return counterexamples before any command reaches the physical layer [13], [16]. This operationalizes the core principle of maintaining guarantees at the control layer by using the digital twin to test LLM proposals against plant dynamics and physical constraints, thereby closing the outer loop with feedback rather than unverified trust. Complementary runtime tooling reinforces this architecture by performing real-time anomaly detection with reactive replanning derived from the LLM's scene assessment [124], while enforcing customizable safety specifications on LLM agents during execution [163].

The second prospective direction focuses on constraint- and safety-preserving supervision. Building on runtime safety filters, CBFs [12], and conformal-prediction guarantees for planners [39], the ultimate objective is supervision that is provably admissible, where the LLM configures references, operational modes, or network topologies while a certified layer guarantees the final execution. Neuro-symbolic approaches that pair LLM reasoning with symbolic constraint solving represent an active research thread toward producing constraint-guaranteed plans [13]. The available toolbox for this architecture is now substantial, incorporating predictive and unified safety filters [36], [37], Hamilton-Jacobi reachability filters [64], formally synthesized neural certificates [66], and conformal Signal Temporal Logic runtime verification [164]. Consequently, the central open challenge is not inventing individual safety mechanisms, but rather assembling them into an integrated, LLM-admitting architecture backed by an end-to-end formal guarantee.

The third prospective direction involves autonomous, tool-using, multi-agent operation. Agentic architectures that integrate perception, decision-making, and action while autonomously calling external tools such as solvers, verifiers, and simulators represent the key trajectory for CNS-scale coordination [13], [16]. Representing external tools as discrete tokens so the model can invoke them natively is one concrete proposal to reduce tool-use overhead [16]. Broader trends across domain surveys highlight cross-domain integration and energy-aware, sustainable system operation [15]. Early evidence of feasibility emerges from LLM-agent control-design frameworks [48] and multi-agent LLM architectures built for 6G network coordination [80], though standardized benchmarks of control competence [33] remain necessary to measure progress. This multi-agent paradigm builds upon a maturing agentic substrate that includes reason-and-act loops [44], tool-learning methodologies [62], and multi-agent coordination frameworks [61], [142], [147], all supported by the broader expansion of foundation models in robotics [151].

The fourth prospective direction explores relaxing the supervisory assumption. The organizing principle of this survey, treating the LLM as a slow supervisor over a fast, certified inner loop, reflects what is currently provable rather than an impassable fundamental barrier, and a growing body of VLA policies [9], [10] already places LLM-derived models directly inside the fast control loop. The key open question is determining what is required to admit an LLM closer to, or inside, the inner loop without sacrificing theoretical guarantees. Three ingredients appear necessary to achieve this: real-time-capable models achieved through distillation and compact architectures to satisfy inner-loop timing; formally verified or filter-wrapped policies so that even an in-loop LLM controller is shadowed by a fast certified safety filter that retains ownership of stability; and a bounded-error characterization of the model itself, without which no closed-loop certificate is possible. This relaxation does not abandon the survey's thesis so much as relocate it, because the formal guarantee still resides in a fast, certified element such as a safety filter rather than a supervisory controller, while the LLM simply operates nearer the actuator.

## 6.4 Conclusion

This survey has taken a unified, network-aware view of LLMs integrated into the loop of networked control, cyber-physical, and multi-agent systems. Throughout this analysis, it has argued a single organizing claim: the LLM is most productively treated as a slow, context-aware supervisory agent that reshapes system objectives, communication, and coordination, while stability and safety are rigorously enforced by the underlying control layer. The unified model of Section 2 makes this relationship precise by framing the LLM as a two-timescale modulation. In doing so, it demonstrates that an LLM in the loop is itself a networked control problem, so classical control tools transfer directly to these architectures.

The empirical evidence assembled across autonomous driving, networking and network operations, and CPSs remains consistent on two fundamental points. First, across every domain, the LLM is consistently assigned to a supervisory role rather than the inner control loop, precisely as the timescale and formal guarantee dimensions of our taxonomy predict. Second, the defining open challenge across the entire field is identical everywhere: establishing a bounded characterization of LLM error and developing a formal stability analysis built upon it. Closing this fundamental gap to convert statistical confidence in LLM outputs into formal closed-loop certificates will ultimately transition LLM-in-the-loop systems from experimental demonstrations to safety-critical deployments.

**Declaration of Generative AI Assistance:** The authors used generative AI tools solely to assist with language editing, formatting, and preliminary analytical drafting during the preparation of this manuscript. The core conceptual framework, survey structure, analysis, and selection of reviewed literature were conceived, conducted, and critically evaluated entirely by the authors. The authors take full responsibility for the contents of the final publication.

# References

[1] J. P. Hespanha, P. Naghshtabrizi, and Y. Xu, "A survey of recent results in networked control systems," Proceedings of the IEEE, vol. 95, no. 1, pp. 138-162, 2007.

[2] X.-M. Zhang, Q.-L. Han, and X. Yu, "Survey on recent advances in networked control systems," IEEE Transactions on Industrial Informatics, vol. 12, no. 5, pp. 1740-1752, 2016.

[3] K.-D. Kim and P. R. Kumar, "Cyber-physical systems: A perspective at the centennial," Proceedings of the IEEE, vol. 100, no. Special Centennial Issue, pp. 1287-1308, 2012.

[4] R. Rajkumar, I. Lee, L. Sha, and J. Stankovic, "Cyber-physical systems: The next computing revolution," Proceedings of the 47th Design Automation Conference (DAC), 2010, pp. 731-736.

[5] R. Olfati-Saber, J. A. Fax, and R. M. Murray, "Consensus and cooperation in networked multi-agent systems," Proceedings of the IEEE, vol. 95, no. 1, pp. 215-233, 2007.

[6] T. B. Brown et al., "Language models are few-shot learners," Advances in Neural Information Processing Systems (NeurIPS), arXiv:2005.14165, 2020.

[7] W. X. Zhao et al., "A survey of large language models," Frontiers of Computer Science, vol. 20, Art. No. 2012627, 2026. https://doi.org/10.1007/s11704-026-60308-3.

[8] B. Ichter et al., "Do as I can, not as I say: Grounding language in robotic affordances," Proceedings of The 7th Conference on Robot Learning, PMLR vol. 205, pp. 287-318, 2023.

[9] B. Zitkovich et al., "RT-2: Vision-language-action models transfer web knowledge to robotic control," Proceedings of The 7th Conference on Robot Learning, PMLR vol. 229, pp.2165-2183, 2023.

[10] M. J. Kim et al., "OpenVLA: An open-source vision-language-action model," arXiv preprint arXiv:2406.09246, 2024.

[11] M. Fakih et al., "LLM4PLC: Harnessing large language models for verifiable programming of PLCs in industrial control systems," Proceedings of the 46th International Conference on Software Engineering: Software Engineering in Practice, 2024, pp. 192-203.

[12] A. D. Ames et al., "Control barrier functions: Theory and applications," 18th European Control Conference, 2019, pp. 3420-3431.

[13] W. Xu, M. Liu, O. Sokolsky, I. Lee, and F. Kong, "LLM-enabled cyber-physical systems: Survey, research opportunities, and challenges," IEEE International Workshop on Foundation Models for Cyber-Physical Systems & Internet of Things, 2024, pp. 50-55.

[14] Y. Zhu, S. Wang, W. Zhong, N. Shen, Y. Li, S. Wang, Z. Li, C. Wu, Z. He, and L. Li, "A survey on large language model-powered autonomous driving," Engineering, 2025, in press, https://doi.org/10.1016/j.eng.2025.07.038.

[15] S. Long, J. Tan, B. Mao, F. Tang, Y. Li, M. Zhao, and N. Kato, "A survey on intelligent network operations and performance optimization based on large language models," IEEE Communications Surveys & Tutorials, vol. 27, no. 6, pp. 3915-3949, 2025.

[16] C. Liu, X. Xie, X. Zhang, and Y. Cui, "Large language models for networking: Workflow, advances, and challenges," IEEE Network, vol. 39, no. 5, pp. 165-172, 2025.

[17] K. Nosrati, A. Tepljakov, J. Belikov, and E. Petlenkov, “When control meets large language models: From words to dynamics,” Engineering Applications of Artificial Intelligence, vol. 178, art. no. 115119, 2026.

[18] L. Ouhib, “Using large language models in control system design: A systematic literature review,” 2026. https://doi.org/10.13140/RG.2.2.35424.85762.

[19] P. Li et al., "Large language models for multi-robot systems: A survey," Autonomous Robots, 50:30, 2026. https://doi.org/10.1007/s10514-026-10257-4.

[20] M. Sarwar, M. Rizwan, M. Aziz, and A. R. Sudais, "Large language models for power system applications: A comprehensive literature survey,"arXiv:2512.13004, 2025.

[21] H. K. Khalil, Nonlinear Systems, 3rd ed. Upper Saddle River, NJ: Prentice Hall, 2002.

[22] E. D. Sontag, "Input to state stability: Basic concepts and results," Nonlinear and Optimal Control Theory, Lecture Notes in Mathematics, vol. 1932. Berlin: Springer, 2008, pp. 163-220.

[23] W. Ren and R. W. Beard, Distributed Consensus in Multi-vehicle Cooperative Control: Theory and Applications. London: Springer, 2008.

[24] W. P. M. H. Heemels, K. H. Johansson, and P. Tabuada, "An introduction to event-triggered and self-triggered control," IEEE 51st IEEE Conference on Decision and Control, 2012, pp. 3270-3285.

[25] L. Huang et al., "A survey on hallucination in large language models: Principles, taxonomy, challenges, and open questions," ACM Transactions on Information Systems, vol. 43, no. 2, pp. 1-55, 2025.

[26] F. L. Lewis, S. Jagannathan, and A. Yeşildirek, Neural Network Control of Robot Manipulators and Nonlinear Systems. London: Taylor & Francis, 1999.

[27] R. S. Sutton and A. G. Barto, Reinforcement Learning: An Introduction, 2nd ed. Cambridge, MA: MIT Press, 2018.

[28] J. García and F. Fernández, "A comprehensive survey on safe reinforcement learning," Journal of Machine Learning Research, vol. 16, no. 42, pp. 1437-1480, 2015.

[29] L. Hewing, K. P. Wabersich, M. Menner, and M. N. Zeilinger, "Learning-based model predictive control: Toward safe learning in control," Annual Review of Control, Robotics, and Autonomous Systems, vol. 3, pp. 269-296, 2020.

[30] L. Brunke, M. Greeff, A. W. Hall, Z. Yuan, S. Zhou, J. Panerati, and A. P. Schoellig, "Safe learning in robotics: From learning-based control to safe reinforcement learning," Annual Review of Control, Robotics, and Autonomous Systems, vol. 5, pp. 411-444, 2022.

[31] D. Driess et al., "PaLM-E: An embodied multimodal language model," Proceedings of the 40th International Conference on Machine Learning, 2023, pp. 8469-8488.

[32] R. Bommasani et al., "On the opportunities and risks of foundation models," arXiv:2108.07258, 2021.

[33] D. Kevian, U. Syed, X. Guo, A. Havens, G. Dullerud, P. Seiler, L. Qin, and B. Hu, "Capabilities of large language models in control engineering: A benchmark study on GPT-4, Claude 3 Opus, and Gemini 1.0 Ultra," arXiv:2404.03647, 2024.

[34] F. Berkenkamp, M. Turchetta, A. P. Schoellig, and A. Krause, "Safe model-based reinforcement learning with stability guarantees," Proceedings of the 31st International Conference on Neural Information Processing Systems, 2017, pp. 908-919.

[35] J. F. Fisac, A. K. Akametalu, M. N. Zeilinger, S. Kaynama, J. Gillula, and C. J. Tomlin, "A general safety framework for learning-based control in uncertain robotic systems," IEEE Transactions on Automatic Control, vol. 64, no. 7, pp. 2737-2752, 2019.

[36] K. P. Wabersich and M. N. Zeilinger, "A predictive safety filter for learning-based control of constrained nonlinear dynamical systems," Automatica, vol. 129, Art. no. 109597, 2021.

[37] K.-C. Hsu, H. Hu, and J. F. Fisac, "The safety filter: A unified view of safety-critical control in autonomous systems," Annual Review of Control, Robotics, and Autonomous Systems, vol. 7, pp. 47-72, 2024.

[38] Y. Chow, O. Nachum, E. Duenez-Guzman, and M. Ghavamzadeh, "A Lyapunov-based approach to safe reinforcement learning," Proceedings of the 32nd International Conference on Neural Information Processing Systems, 2018, pp. 8103-8112.

[39] A. Z. Ren et al., "Robots that ask for help: Uncertainty alignment for large language model planners," Proceedings of The 7th Conference on Robot Learning, PMLR vol. 229, 2023, pp. 661-682.

[40] S. Hong et al., "MetaGPT: Meta programming for a multi-agent collaborative framework," arXiv:2308.00352, 2023.

[41] Q. Wu, G. Bansal, J. Zhang, Y. Wu, B. Li, E. Zhu, et al., "AutoGen: Enabling next-gen LLM applications via multi-agent conversation," arXiv:2308.08155, 2023.

[42] Z. Mandi, S. Jain, and S. Song, "RoCo: Dialectic multi-robot collaboration with large language models," IEEE International Conference on Robotics and Automation, 2024, pp. 286-299.

[43] B. Liu, Y. Jiang, X. Zhang, Q. Liu, S. Zhang, J. Biswas, and P. Stone, "LLM+P: Empowering large language models with optimal planning proficiency," arXiv:2304.11477, 2023.

[44] S. Yao, J. Zhao, D. Yu, N. Du, I. Shafran, K. Narasimhan, and Y. Cao, "ReAct: Synergizing reasoning and acting in language models," arXiv:2210.03629, 2022.

[45] W. Huang, P. Abbeel, D. Pathak, and I. Mordatch, "Language models as zero-shot planners: Extracting actionable knowledge for embodied agents," Proceedings of the 39th International Conference on Machine Learning, PMLR, VOL. 162, pp. 9118-9147, 2022.

[46] Y. Xie, C. Yu, T. Zhu, J. Bai, Z. Gong, and H. Soh, "Translating natural language to planning goals with large-language models," arXiv:2302.05128, 2023.

[47] S. K. Jha et al., "Neuro-symbolic reasoning for planning: Counterexample-guided inductive synthesis using large language models and satisfiability solving," arXiv:2309.16436, 2023.

[48] X. Guo, D. Keivan, U. Syed, L. Qin, H. Zhang, G. Dullerud, P. Seiler, and B. Hu, "ControlAgent: Automating control system design via novel integration of LLM agents and domain expertise," arXiv:2410.19811, 2024.

[49] K. Tohma, H. İ. Okur, H. Gürsoy-Demir, M. N. Aydın, and C. Yeroğlu, "SmartControl: Interactive PID controller design powered by LLM agents and control system expertise," SoftwareX, vol. 31, Art. no. 102194, 2025.

[50] Y. J. Ma, W. Liang, G. Wang, D.-A. Huang, O. Bastani, D. Jayaraman, Y. Zhu, L. Fan, and A. Anandkumar, "Eureka: Human-level reward design via coding large language models," arXiv:2310.12931, 2023.

[51] T. Xie, S. Zhao, C. H. Wu, Y. Liu, Q. Luo, V. Zhong, Y. Yang, and T. Yu, "Text2Reward: Reward shaping with language models for reinforcement learning," arXiv:2309.11489, 2023.

[52] W. Yu et al., "Language to rewards for robotic skill synthesis," arXiv:2306.08647, 2023.

[53] J. Pan, G. Chou, and D. Berenson, "Data-efficient learning of natural language to linear temporal logic translators for robot task specification," IEEE International Conference on Robotics and Automation, 2023, pp. 11554-11561.

[54] D. Mendoza, C. Hahn, and C. Trippel, "Translating natural language to temporal logics with large language models and model checkers," 2024 Formal Methods in Computer-Aided Design, 2024, pp. 119-129.

[55] Z. Liu et al., "Agents4PLC: Automating closed-loop PLC code generation and verification in industrial control systems using LLM-based agents," IEEE Transactions on Software Engineering, vol. 52, no. 5, pp. 1672-1687, 2026.

[56] H. Koziolek et al., "Spec2Control: Automating PLC/DCS control-logic engineering from natural language requirements with LLMs — a multi-plant evaluation," arXiv:2510.04519, 2025.

[57] N. Shinn, F. Cassano, E. Berman, A. Gopinath, K. Narasimhan, and S. Yao, "Reflexion: Language agents with verbal reinforcement learning," Proceedings of the 37th International Conference on Neural Information Processing System, 2023, pp. 8634-8652.

[58] G. Wang, Y. Xie, Y. Jiang, A. Mandlekar, C. Xiao, Y. Zhu, L. Fan, and A. Anandkumar, "Voyager: An open-ended embodied agent with large language models," arXiv:2305.16291, 2023.

[59] L. Song, C. Zhang, L. Zhao, and J. Bian, "Pre-trained large language models for industrial control," arXiv:2308.03028, 2023.

[60] J. S. Park, J. C. O'Brien, C. J. Cai, M. R. Morris, P. Liang, and M. S. Bernstein, "Generative agents: Interactive simulacra of human behavior," Proceedings of the 36th Annual ACM Symposium on User Interface Software and Technology, 2023, pp. 1-22.

[61] C. Sun, S. Huang, and D. Pompili, "LLM-based multi-agent decision-making: Challenges and future directions," IEEE Robotics and Automation Letters, vol. 10, no. 6, pp. 5681-5688, 2025.

[62] Y. Qin et al., "Tool learning with foundation models," ACM Computing Surveys, vol. 57, no. 4, pp. 1-40, 2024.

[63] Z. Yang, S. S. Raman, A. Shah, and S. Tellex, "Plug in the safety chip: Enforcing constraints for LLM-driven robot agents," 2024 IEEE International Conference on Robotics and Automation, 2024, pp. 14435-14442.

[64] J. Borquez, K. Chakraborty, H. Wang, and S. Bansal, "On safety and liveness filtering using Hamilton–Jacobi reachability analysis," IEEE Transactions on Robotics, vol. 40, pp. 4235-4251, 2024.

[65] C. Dawson, S. Gao, and C. Fan, "Safe control with learned certificates: A survey of neural Lyapunov, barrier, and contraction methods for robotics and control," IEEE Transactions on Robotics, vol. 39, no. 3, pp. 1749-1767, 2023.

[66] A. Peruffo, D. Ahmed, and A. Abate, "Automated and formal synthesis of neural barrier certificates for dynamical models," Tools and Algorithms for the Construction and Analysis of Systems (TACAS), 2021, pp. 370-388.

[67] Z. He et al., "Designing network algorithms via large language models," Proceedings of the 23rd ACM Workshop on Hot Topics in Networks, 2024, pp. 205-212.

[68] R. Mondal et al., "What do LLMs need to synthesize correct router configurations?" Proceedings of the 22nd ACM Workshop on Hot Topics in Networks, 2023, pp. 189-195.

[69] X. Lian et al., "Configuration validation with large language models," arXiv:2310.09690, 2023.

[70] S. K. Mani et al., "Enhancing network management using code generated by large language models," Proceedings of the 22nd ACM Workshop on Hot Topics in Networks, 2023, pp. 196-204.

[71] D. Wu et al., "NetLLM: Adapting large language models for networking," Proceedings of the ACM SIGCOMM 2024 Conference, 2024, pp. 661-678.

[72] R. Meng et al., "Large language model guided protocol fuzzing," Proceedings of the 31st Annual Network and Distributed System Security Symposium, 2024. DOI:10.14722/ndss.2024.24556.

[73] A. Mekrache, A. Ksentini, and C. Verikoukis, "Intent-based management of next-generation networks: An LLM-centric approach," IEEE Network, vol. 38, no. 5, pp. 29-36, 2024.

[74] G. Sun et al., "Large language model (LLM)-enabled graphs in dynamic networking," IEEE Network, vol. 39, no. 4, pp. 290-301, 2025.

[75] Y. Wang et al., "AlarmGPT: An intelligent alarm analyzer for optical networks using a generative pre-trained transformer," Journal of Optical Communications and Networking, vol. 16, no. 6, pp. 681-694, 2024.

[76] K. B. Kan, H. Mun, G. Cao, and Y. Lee, "Mobile-LLaMA: Instruction fine-tuning open-source LLM for network analysis in 5G networks," IEEE Network, vol. 38, no. 5, pp. 76-83, 2024.

[77] M. Xu et al., "When large language model agents meet 6G networks: Perception, grounding, and alignment," IEEE Wireless Communications, vol. 31, no. 6, pp. 63-71, 2024.

[78] H. Zhou et al., "Large language model (LLM) for telecommunications: A comprehensive survey on principles, key techniques, and opportunities," IEEE Communications Surveys & Tutorials, vol. 27, no. 3, pp. 1955-2005, 2025.

[79] J. Shao, J. Tong, Q. Wu, W. Guo, Z. Li, Z. Lin, and J. Zhang, "WirelessLLM: Empowering large language models towards wireless intelligence," Journal of Communications and Information Networks, vol. 9, no. 2, pp. 99-112, 2024.

[80] F. Jiang, Y. Peng, L. Dong, K. Wang, K. Yang, C. Pan, D. Niyato, and O. A. Dobre, "Large language model enhanced multi-agent systems for 6G communications," IEEE Wireless Communications, vol. 31, no. 6, pp. 48-55, 2024.

[81] J. Liu, C. Zhang, J. Qian, M. Ma, S. Qin, C. Bansal, Q. Lin, S. Rajmohan, and D. Zhang, "Large language models can deliver accurate and interpretable time series anomaly detection," Proceedings of the 31st ACM SIGKDD Conference on Knowledge Discovery and Data Mining V.2, 2024, pp. 4623-4634.

[82] W. Guan, J. Cao, S. Qian, J. Gao, and C. Ouyang, "LogLLM: Log-based anomaly detection using large language models," arXiv:2411.08561, 2024.

[83] M. Guastalla, Y. Li, A. Hekmati, and B. Krishnamachari, "Application of large language models to DDoS attack detection," Security and Privacy in Cyber-Physical Systems and Smart Vehicles, 2024, pp. 83-99.

[84] T. Wang et al., "ShieldGPT: An LLM-based framework for DDoS mitigation," Proceedings of the 8th Asia-Pacific Workshop on Networking, 2024, pp. 108-114.

[85] H. Yang, K. Xiang, M. Ge, H. Li, R. Lu, and S. Yu, "A comprehensive overview of backdoor attacks in large language models within communication networks," IEEE Network, vol. 38, no. 6, pp. 211-218, 2024.

[86] C. M. Ahmed, "AttackLLM: LLM-based attack pattern generation for an industrial control system," arXiv:2504.04187, 2025.

[87] V. Raval, M. Zeid, and P. Enjeti, "Circuit-AI: A self-hosted AI-agent language model framework for control loop implementation and simulation," 2025 IEEE International Telecommunications Energy Conference, 2025, pp. 91-96.

[88] K. Rana, J. Haviland, S. Garg, J. Abou-Chakra, I. Reid, and N. Suenderhauf, "SayPlan: Grounding large language models using 3D scene graphs for scalable robot task planning," Proceedings of The 7th Conference on Robot Learning, PMLR, vol. 229, 2023, pp. 23-72.

[89] W. Huang et al., "Grounded decoding: Guiding text generation with grounded models for robot control," arXiv:2303.00855, 2023.

[90] K. Lin, C. Agia, T. Migimatsu, M. Pavone, and J. Bohg, "Text2Motion: From natural language instructions to feasible plans," Autonomous Robots, vol. 47, pp. 1345-1365, 2023.

[91] W. Huang, C. Wang, R. Zhang, Y. Li, J. Wu, and L. Fei-Fei, "VoxPoser: Composable 3D value maps for robotic manipulation with language models," arXiv:2307.05973, 2023.

[92] W. Huang, C. Wang, Y. Li, R. Zhang, and L. Fei-Fei, "ReKep: Spatio-temporal reasoning of relational keypoint constraints for robotic manipulation," arXiv:2409.01652, 2024.

[93] W. Huang et al., "Inner monologue: Embodied reasoning through planning with language models," Proceedings of The 6th Conference on Robot Learning, PMLR, vol. 205, pp. 1769-1782, 2023.

[94] K. Black et al., "π0: A vision-language-action flow model for general robot control," arXiv:2410.24164, 2024.

[95] X. Li et al., "Vision-language foundation models as effective robot imitators," arXiv:2311.01378, 2023.

[96] J. Liang, W. Huang, F. Xia, P. Xu, K. Hausman, B. Ichter, P. Florence, and A. Zeng, "Code as policies: Language model programs for embodied control," 2023 IEEE International Conference on Robotics and Automation, 2023, pp. 9493-9500.

[97] G. Yoon, S. Lee, and J. Y. Sim, "ModuLoop: Low-level code generation using modular synthesizer and closed-loop debugger for robotic control," IEEE Robotics and Automation Letters, vol. 10, no. 12, pp. 12812-12819, 2025.

[98] E. Vorobiov, A. J. Mahmood, S. Rezvani, and R. Chhabra, "ARRC: Advanced reasoning robot control - Knowledge-driven autonomous manipulation using retrieval-augmented generation," arXiv:2510.05547, 2025.

[99] J. N. Ramos-Silva and P. J. Burke, "A universal large language model-drone command and control interface," arXiv:2601.15486, 2026.

[100] N. Abd Manap, T. C. Yang, and A. Putra, "Embedded voice-controlled AI assistant for robotic arm operation in industrial automation," Journal of Telecommunication, Electronic and Computer Engineering, vol. 17, no. 4, pp. 7-14, 2025.

[101] H. Pan et al., "Robot navigation via foundation language models: A review," ACM Computing Surveys, vol. 58, no. 11, pp. 1-38, 2026.

[102] Z. Xu et al., "DriveGPT4: Interpretable end-to-end autonomous driving via large language model," IEEE Robotics and Automation Letters, vol. 9, no. 10, pp. 8186-8193, 2024.

[103] L. Wen et al., "DiLu: A knowledge-driven approach to autonomous driving with large language models," arXiv:2309.16292, 2023.

[104] X. Tian et al., "DriveVLM: The convergence of autonomous driving and large vision-language models," arXiv:2402.12289, 2024.

[105] E. Cui et al., "DriveMLM: Aligning multi-modal large language models with behavioral planning states for autonomous driving," Visual Intelligence, vol. 3, art. no. 22, 2025.

[106] J. Mao, Y. Qian, J. Ye, H. Zhao, and Y. Wang, "GPT-driver: Learning to drive with GPT," arXiv:2310.01415, 2023.

[107] H. Shao, Y. Hu, L. Wang, S. L. Waslander, Y. Liu, and H. Li, "LMDrive: Closed-loop end-to-end driving with large language models," IEEE/CVF Conference on Computer Vision and Pattern Recognition, 2024, pp. 15120-15130.

[108] C. Sima et al., "DriveLM: Driving with graph visual question answering," European Conference on Computer Vision, 2024, pp. 256-274.

[109] C. Cui, Z. Yang, Y. Zhou, Y. Ma, J. Lu, and Z. Wang, "Large language models for autonomous driving: Real-world experiments," arXiv:2312.09397, 2023.

[110] H. Sha et al., "LanguageMPC: Large language models as decision makers for autonomous driving," arXiv:2310.03026, 2023.

[111] I. de Zarzà, J. de Curtò, G. Roig, and C. T. Calafate, "LLM adaptive PID control for B5G truck platooning systems," Sensors, vol. 23, no. 13, art. no. 5899, 2023.

[112] Y. Zheng et al., "PlanAgent: A multi-modal large language agent for closed-loop vehicle motion planning," arXiv:2406.01587, 2024.

[113] Y. Wang et al., "Empowering autonomous driving with large language models: A safety perspective," arXiv:2312.00812, 2024.

[114] J. Chen, S. Dai, F. Chen, Z. Lv, J. Tang, and L. Han, "Edge-cloud collaborative motion planning for autonomous driving with large language models," IEEE 24th International Conference on Communication Technology, 2024, pp. 185-190.

[115] K. Long, H. Shi, J. Liu, and X. Li, "VLM-MPC: Vision language foundation model (VLM)-guided model predictive controller (MPC) for autonomous driving," arXiv:2408.04821, 2024.

[116] T. Zeng and Y. Zhang, "Large language model-augmented model predictive control for marine vessels in uncertain marine environments," Ocean Engineering, vol. 346, art. no. 123628, 2026.

[117] Y. Xia, M. Shenoy, N. Jazdi, and M. Weyrich, "Towards autonomous system: Flexible modular production system enhanced with large language model agents," IEEE 28th International Conference on Emerging Technologies and Factory Automation, 2023. DOI: 10.1109/ETFA54631.2023.10275362.

[118] N. Yoshikawa et al., "Large language models for chemistry robotics," Autonomous Robots, vol. 47, pp. 1057-1086, 2023.

[119] D. A. Boiko, R. MacKnight, B. Kline, and G. Gomes, "Autonomous chemical research with large language models," Nature, vol. 624, pp. 570-578, 2023.

[120] H. Cui, Y. Du, Q. Yang, Y. Shao, and S. C. Liew, "LLMind: Orchestrating AI and IoT with LLMs for complex task execution," IEEE Communications Magazine, vol. 63, no. 4, pp. 214-220, 2025.

[121] A. Rasheed, O. Ravik, and O. San, "Large language models for control," arXiv:2511.00337, 2025.

[122] A. E. Ojuolape and S. Hu, "Explainable fault diagnosis of control systems using large language models," IEEE Conference on Control Technology and Applications, 2024, pp. 491-498.

[123] L. Tao, H. Liu, G. Ning, W. Cao, B. Huang, and C. Lu, "LLM-based framework for bearing fault diagnosis," Mechanical Systems and Signal Processing, vol. 224, art. no. 112127, 2025.

[124] R. Sinha, A. Elhafsi, C. Agia, M. Foutter, E. Schmerling, and M. Pavone, "Real-time anomaly detection and reactive planning with large language models," arXiv:2407.08735, 2024.

[125] Ł. Żukowski and J. F. Mozaryn, "Evaluating large language models for graphical PLC programming: A temperature control case study," 29th International Conference on Methods and Models in Automation and Robotics, 2025, pp. 339-344.

[126] Y. Xia, J. Zhang, N. Jazdi, and M. Weyrich, "Incorporating large language models into production systems for enhanced task automation and flexibility," arXiv:2407.08550, 2024.

[127] Y. Zhang, A. M. Saber, A. Youssef, and D. Kundur, "Grid-Agent: An LLM-powered multi-agent system for power grid control," arXiv:2508.05702, 2025.

[128] A. Jena, F. Ding, J. Wang, Y. Yao, and L. Xie, "LLM-based adaptive distribution voltage regulation under frequent topology changes: An in-context MPC framework," IEEE Transactions on Smart Grid, vol. 16, no. 5, pp. 4297-4300, 2025.

[129] F. Lin, X. Li, W. Lei, J. J. Rodriguez-Andina, J. M. Guerrero, and C. Wen, "PE-GPT: A new paradigm for power electronics design," IEEE Transactions on Industrial Electronics, vol. 72, no. 4, pp. 3778-3791, 2024.

[130] S. Lai, Z. Xu, W. Zhang, H. Liu, and H. Xiong, "LLMLight: Large language models as traffic signal control agents," arXiv:2312.16044, 2023.

[131] M. Narimani and S. A. Emami, "AgenticControl: An automated control design framework using large language models," arXiv:2506.19160, 2025.

[132] R. Zahedifar, S. A. Mirghasemi, M. S. Baghshah, and A. Taheri, "LLM-Agent-Controller: A universal multi-agent large language model system as a control engineer," arXiv:2505.19567, 2025.

[133] K. Gashi, M. Chacin, and G. Tang, "Electromechanical design and control using a large language model," 12th International Conference on Control, Mechatronics and Automation, 2024, pp. 354-360.

[134] A. A. Khan, M. Andrev, M. A. Murtaza, S. Aguilera, R. Zhang, J. Ding, S. Hutchinson, and A. Anwar, "Safety-aware task planning via large language models in robotics," IEEE/RSJ International Conference on Intelligent Robots and Systems, 2025, pp. 21024-21031.

[135] L. Brunke, Y. Zhang, R. Römer, J. Naimer, N. Staykov, S. Zhou, and A. P. Schoellig, "Semantically safe robot manipulation: From semantic scene understanding to motion safeguards," IEEE Robotics and Automation Letters, vol. 10, no. 5, pp. 4810-4817, 2025.

[136] N. Nascimento, P. Alencar, and D. Cowan, "GPT-in-the-loop: Adaptive decision-making for multi-agent systems," arXiv:2308.10435, 2023.

[137] Z. Wang et al., "RALLY: Role-adaptive LLM-driven yoked navigation for agentic UAV swarms," IEEE Open Journal of Vehicular Technology, vol. 6, pp. 2693-2708, 2025.

[138] M. Schuck, D. O. Dahanaggamaarachchi, B. Sprenger, V. Vyas, S. Zhou, and A. P. Schoellig, "SwarmGPT: Combining large language models with safe motion planning for drone swarm choreography," IEEE Robotics and Automation Letters, vol. 10, no. 11, pp. 12237-12244, 2025.

[139] A. Shibu, M. Saleh, M. Al-Musleh, and N. Abdulaziz, "SkySim: A ROS2-based simulation environment for natural language control of drone swarms using large language models," arXiv:2602.01226, 2026.

[140] A. Ahmed, L. Wang, J. Kim, J. Jin, K. Cho, C. Kwon, and D. J. Lee, "LLM-guided distributed model predictive control for decentralized UAV formations," IEEE Access, vol. 14, pp. 15226-15240, 2025.

[141] Y. Wu et al., "Multi-agent autonomous driving systems with large language models: A survey of recent advances," arXiv:2502.16804, 2025.

[142] H. Zhang, W. Du, J. Shan, Q. Zhou, Y. Du, J. B. Tenenbaum, T. Shu, and C. Gan, "Building cooperative embodied agents modularly with large language models," arXiv:2307.02485, 2023.

[143] J. Chen et al., "EMOS: Embodiment-aware heterogeneous multi-robot operating system with LLM agents," arXiv:2410.22662, 2024.

[144] K. Liu, Z. Tang, D. Wang, Z. Wang, X. Li, and B. Zhao, "COHERENT: Collaboration of heterogeneous multi-robot system with large language models," 2025 IEEE International Conference on Robotics and Automation, 2025, pp. 10208-10214.

[145] B. Zhang et al., "Controlling large language model-based agents for large-scale decision-making: An actor-critic approach," arXiv:2311.13884, 2023.

[146] K. Garg, J. Arkin, S. Zhang, N. Roy, and C. Fan, "Large language models to the rescue: Deadlock resolution in multi-robot systems," arXiv:2404.06413, 2024.

[147] S. Hu, Z. Fang, Z. Fang, Y. Deng, X. Chen, and Y. Fang, "AgentsCoDriver: Large language model empowered collaborative driving with lifelong learning," arXiv:2404.06345, 2024.

[148] S. Fang, J. Liu, Y. Cui, C. Lv, P. Hang, and J. Sun, "Towards interactive and learnable cooperative driving automation: A large language model-driven decision-making framework," IEEE Transactions on Vehicular Technology, vol. 74, no. 8, pp. 11894-11905, 2025.

[149] Z. Yuan, S. Lai, and H. Liu, "CoLLMLight: Cooperative large language model agents for network-wide traffic signal control," arXiv:2503.11739, 2025.

[150] A. Robey, Z. Ravichandran, V. Kumar, H. Hassani, and G. J. Pappas, "Jailbreaking LLM-controlled robots," IEEE International Conference on Robotics and Automation, 2025, pp. 11948-11956.

[151] R. Firoozi et al., "Foundation models in robotics: Applications, challenges, and the future," International Journal of Robotics Research, vol. 44, no. 5, pp. 701-739, 2025.

[152] M. Shridhar et al., "ALFRED: A benchmark for interpreting grounded instructions for everyday tasks," IEEE/CVF Conference on Computer Vision and Pattern Recognition, 2020, pp. 10737-10746.

[153] S. Srivastava et al., "BEHAVIOR: Benchmark for everyday household activities in virtual, interactive, and ecological environments," Proceedings of the 5th Conference on Robot Learning, PMLR, vol. 164, 2022, pp. 477-490.

[154] Z. Zhou et al., "A survey on efficient inference for large language models," arXiv:2404.14294, 2024.

[155] Gemma Team, "Gemma: Open models based on Gemini research and technology," arXiv:2403.08295, 2024.

[156] Y. Liu et al., "Hallucination-aware optimization for large language model-empowered communications," IEEE Communications Magazine, vol. 64, no. 3, pp. 24-31, 2026.

[157] G. Katz, C. Barrett, D. L. Dill, K. Julian, and M. J. Kochenderfer, "Reluplex: An efficient SMT solver for verifying deep neural networks," International Conference on Computer Aided Verification, 2017, pp. 97-117.

[158] R. Ivanov, J. Weimer, R. Alur, G. J. Pappas, and I. Lee, "Verisig: Verifying safety properties of hybrid systems with neural network controllers," Proceedings of the 22nd ACM International Conference on Hybrid Systems: Computation and Control, 2019, pp. 169-178.

[159] D. Zhi, P. Wang, S. Liu, C.-H. L. Ong, and M. Zhang, "Unifying qualitative and quantitative safety verification of DNN-controlled systems," International Conference on Computer Aided Verification, 2024, pp. 401-426.

[160] Y. Yao, J. Duan, K. Xu, Y. Cai, Z. Sun, and Y. Zhang, "A survey on large language model (LLM) security and privacy: The good, the bad, and the ugly," High-Confidence Computing, vol. 4, no. 2, art. no. 100211, 2024.

[161] H. Zhang et al., "BadRobot: Jailbreaking embodied LLMs in the physical world," arXiv:2407.20242, 2024.

[162] Z. Chen, Z. Xiang, C. Xiao, D. Song, and B. Li, "AgentPoison: Red-teaming LLM agents via poisoning memory or knowledge bases," Proceedings of the 38th International Conference on Neural Information Processing System, 2024, pp. 130185-130213.

[163] H. Wang, C. M. Poskitt, and J. Sun, "AgentSpec: Customizable runtime enforcement for safe and reliable LLM agents," arXiv:2503.18666, 2025.

[164] L. Lindemann, X. Qin, J. V. Deshmukh, and G. J. Pappas, "Conformal prediction for STL runtime verification," Proceedings of the ACM/IEEE 14th International Conference on Cyber-Physical Systems (with CPS-IoT Week 2023), 2023, pp. 142-153.